\documentclass{aa}  
\usepackage{graphicx}
\usepackage{natbib}
\bibpunct{(}{)}{:}{a}{}{,}
\usepackage{txfonts}
\begin{document}
   \title{Line-profile variations of stochastically excited oscillations in four evolved stars\thanks{The software package FAMIAS developed in the framework of the FP6 European Coordination Action HELAS (http://www.helas-eu.org) has been used in this research.}}

   \author{S. Hekker \inst{1} \fnmsep \inst{2} \fnmsep \inst{3}
    \and C. Aerts \inst{2} \fnmsep \inst{4}}

   \offprints{S. Hekker, \\
                    email: saskia@bison.ph.bham.ac.uk}
   
   \institute{School of Physics and Astronomy, University of Birmingham, Edgbaston, Birmingham B15 2TT, United Kingdom
                  \and
              Instituut voor Sterrenkunde, Katholieke Universiteit Leuven,
Celestijnenlaan 200 D, 3001 Leuven, Belgium     
                  \and
              Royal Observatory of Belgium, Ringlaan 3, 1180 Brussels, Belgium
                    \and
              Department of Astrophysics, IMAP, University of Nijmegen, PO Box 9010,
6500 GL Nijmegen, The Netherlands }

   \date{Received <date>; accepted <date>}

 
  \abstract
  {Since solar-like oscillations were first detected in red-giant stars, the presence of non-radial oscillation modes has been debated. Spectroscopic line-profile analysis was used in the first attempt to perform mode identification, which revealed that non-radial modes are observable. Despite the fact that the presence of non-radial modes could be confirmed, the degree or azimuthal order could not be uniquely identified. Here we present an improvement to this first spectroscopic line-profile analysis.}
{We aim to study line-profile variations of stochastically excited solar-like oscillations in four evolved stars to derive the azimuthal order of the observed mode and the surface rotational frequency.}
{Spectroscopic line-profile analysis is applied to cross-correlation functions, using the Fourier Parameter Fit method on the amplitude and phase distributions across the profiles.}
{For four evolved stars, $\beta$~Hydri (G2IV), $\epsilon$~Ophiuchi (G9.5III), $\eta$~Serpentis (K0III) and $\delta$~Eridani (K0IV) the line-profile variations reveal the azimuthal order of the oscillations with an accuracy of $\pm$~1. Furthermore, our analysis reveals the projected rotational velocity and the inclination angle. From these parameters we obtain the surface rotational frequency.}
 {We conclude that line-profile variations of cross-correlation functions behave differently for different frequencies and that they provide additional information in terms of the surface rotational frequency and azimuthal order.}

 \authorrunning{S. Hekker \& C. Aerts}
  \maketitle
%

\section{Introduction}
After several previous claims, the first firm observational evidence of solar-like oscillations in red-giant stars was presented by \citet{frandsen2002} for $\xi$~Hydrae. This was followed by discoveries from two-site ground-based spectroscopic campaigns on $\epsilon$~Ophiuchi \citep{deridder2006} and $\eta$~Serpentis \citep{barban2004}. Based on theory of more luminous red giants \citep{dziembowski2001}, the detected frequencies of these stars were interpreted as radial modes and the stars were modelled based on this assumption \citep{houdek2002,deridder2006}.

\citet{hekker2006} made the first time-series analysis of spectral line shape variations and attempted to perform spectroscopic mode identification of the observed frequencies for three red-giant stars ($\epsilon$~Ophiuchi, $\eta$~Serpentis and $\xi$~Hydrae) and one subgiant ($\delta$~Eridani). They investigated line-profile variations in the cross-correlation functions based on a pixel-by-pixel method, i.e., fitting a sinusoid at every velocity pixel across the profile. From the differences in the shape of the amplitude distribution across the profile for different frequencies they concluded that non-radial oscillations have to be present, although no definite identification could be provided.

The CoRoT (Convection Rotation and planetary Transits) satellite performs photometry on parts of the sky for 150 consecutive days with 32 / 512 second cadences in the centre and anti-centre direction of the galaxy. Among the observed stars are many red giants. With these observations the existence of non-radial oscillations in red-giant stars has been firmly proven \citep{deridder2009}. This proof is based on the determination of the harmonic degrees, using the asymptotic relation for high-order low-degree modes \citep{tassoul1980}. This relation predicts that non-radial modes appear at regular intervals in the Fourier spectrum (large separation), with modes with harmonic degree $\ell$ = 1 approximately half way in between the radial modes and modes with harmonic degree $\ell$ = 2 close to the radial modes (small separation). Indeed, multiple 'ridges' are present in \'{e}chelle diagrams (frequency vs. frequency modulo large separation) of several stars presented by \citet{deridder2009}. These results are confirmed by observations of red giants from the NASA Kepler satelite, see  \citet{bedding2010}, and will be followed up.

\begin{table*}
\begin{minipage}{17cm}
\caption{Stellar parameters of $\epsilon$~Ophiuchi, $\eta$~Serpentis, $\delta$~Eridani and $\beta$~Hydri: Effective temperature ($T_{\mathrm{eff}}$) in Kelvin, radius ($R$) in R$_{\sun}$, mass ($M$) in M$_{\sun}$, metallicity ($\mathrm{[Fe}/\mathrm{H]}$), surface gravity ($\log$ g) in c.g.s units, rotational velocity ($\upsilon \mathrm{sin}i$), microturbulence ($\xi_{\rm micro}$) and macroturbulence ($\xi_{\rm macro}$) in km\,s$^{-1}$, parallax ($\pi$) in mas and the apparent magnitude in the V band ($m_{v}$).}
\label{propstar}
\centering
\renewcommand{\footnoterule}{}
\begin{tabular}{lcccc}
\hline\hline
parameter & $\epsilon$~Ophiuchi & $\eta$~Serpentis  & $\delta$~Eridani & $\beta$~Hydri\\
\hline
$T_{\mathrm{eff}}$ [K] & $4970 \pm 80$\footnote{\citet{hekker2007}, $^b$\citet{richichi2005},  $^{c}$~\citet{kallinger2008}, $^d$~\citet{esa1997}, $^e$~\citet{reiners2003}, $^f$~\citet{valenti2005}, $^g$~\citet{valenti2005} isochrone stellar mass, $^h$~\citet{santos2004}, $^i$~\citet{thevenin2005}, $^j$~\citet{north2007}, $^k$~\citet{dravins1998}} & $4955 \pm 80^a$ & $5050\pm100^h$ & $5872 \pm 44^j$\\
$R$ [R$_{\sun}$] & $10.4 \pm 0.45^{b+d,c}$ & $5.914 \pm 0.094^{b+d}$ or $5.794 \pm 0.097^f$ & $2.33 \pm 0.03^i$ & $1.814 \pm 0.017^j$\\
$M$ [M$_{\sun}$] & $2.0^{c}$ & $2.42 \pm 0.2^f$ or $1.72^{+0.15\,g}_{-0.11}$ & $1.1215^i$ & $1.07 \pm 0.03^j$\\
$\mathrm{[Fe}/\mathrm{H]}$ & $-0.07 \pm 0.11^{a}$ & $-0.15 \pm 0.11^{a}$ & $0.13 \pm 0.03^h$ & $-0.2^k$ \\
$\log$ g (c.g.s.) & $2.9 \pm 0.15^{a}$ & $3.2 \pm 0.15^a$ & $3.77 \pm 0.16^h$ & $3.952 \pm 0.005^j$\\
$\upsilon \mathrm{sin}i$ [km\,s$^{-1}$] & $3.5 \pm 0.5^{a}$ & $0.44 \pm 0.44^a$  & $2.3\pm0.5^e$ & $3.3 \pm 0.3^e$\\
$\xi_{\rm micro}$ [km\,s$^{-1}$]  & $1.5 \pm 0.1^{a}$ & $1.3 \pm 0.1^{a}$ & & \\
$\xi_{\rm macro}$ [km\,s$^{-1}$] & $3.55 \pm 0.45^a$ & $3.52 \pm 0.45^a$ & & \\
$\pi$ [mas] & $30.34\pm0.79^d$ & $52.81\pm0.75^d$ &  $110.58\pm0.88^d$ & $133.78 \pm 0.5^d$ \\
$m_{v}$ [mag] & $3.24\pm0.02^d$ & $3.23\pm0.02^d$ & $3.52\pm0.02^d$ & $2.82 \pm 0.02^d$\\
\hline
\end{tabular}
\end{minipage}
\end{table*}

The asymptotic approximation is valid for p modes and can only be applied for giants and subgiants when the oscillation modes are trapped in the outer parts of the star. The trapping depends on the internal stellar structure. For stars in certain evolutionary states, it is inefficient, such that g modes and mixed modes are also observable, see e.g. \citet{dupret2009}. This degrades the regular pattern, and the asymptotic relation is less useful as a guide to perform identification of the mode degree $\ell$.

For non-radial solar-like oscillations it is expected that all azimuthal orders are excited. However, it is usually not possible to distinguish between oscillation frequencies of modes with the same radial order ($n$), harmonic degree ($\ell$), and different azimuthal order ($m$), due to the generally slow rotational velocity of evolved solar-like oscillators, the limited time span of observations, and stochastic nature of the oscillations. Slow rotation induces only a very small frequency splitting, while the stochastic nature causes the oscillation frequencies to appear as Lorentzian profiles in the Fourier spectrum with a width depending on the lifetime of the mode, i.e., shorter mode lifetimes imply broader profiles. We need long time series of data to obtain sub-$\mu$Hz frequency resolution and to resolve the modes, which can have lifetimes of the order of a few to hundreds of days \citep{dupret2009}.  So, the frequency of a non-radial oscillation mode is extracted from a broad profile consisting of several modes with different $m$ values. Which of these modes is dominant depends on the inclination angle. In stars seen at high inclination angle $m$ $\neq$ 0 modes have higher visibility than $m$ = 0 modes \citep{gizon2003}. Furthermore, for individual realisations, as we consider a single time series of observations, the stochastic interference with noise might boost or diminish a mode. Therefore, we are in principle able to observe any of the present azimuthal orders or combinations thereof. The latter will also behave different  than $m$ = 0 modes due to the influence of $m$ $\neq$ 0 modes.

Line-profile variations are mainly sensitive to the azimuthal order of a mode and much less sensitive to its harmonic degree \citep[see e.g.][]{aerts2000}. The determination of a non-zero azimuthal order would indicate that the oscillation mode is non-radial. This can be important in cases where the asymptotic approximation fails or provides ambiguous results, i.e., where different ridges in an \'{e}chelle diagram can not unambiguously be identified. In case $\ell$ is known from the asymptotic approximation the $m$ value can put additional constraints on the internal structure. Also, we can obtain the inclination angle for non-radial modes and the projected rotational velocity from line-profile analysis. These parameters allow to determine the stellar rotational frequency for non-radial oscillators.

With this potential of the analysis of line-profile variations in mind, we have improved the spectroscopic line-profile analysis originally presented by \citet{hekker2006} (Section 3). These improvements were necessary as \citet{hekker2006} were only able to match observations and theory qualitatively. In this work we perform a quantitative analysis. We use FAMIAS \citep[Frequency Analysis and Mode Identification for Asteroseismology,][]{zima2008}, a package of state-of-the art tools for the analysis of photometric and spectroscopic time-series data, to perform line-profile analysis (Section 4) on three stars, $\beta$~Hydri (G2IV) $\delta$ Eridani (K0IV) and $\epsilon$~Ophiuchi (G9.5III), for which previous mode identification was available \citep{bedding2007,carrier2002,carrier2003,kallinger2008}.

In Section 5, we analyse the observed line-profile variations for oscillation modes of $\beta$~Hydri, $\delta$ Eridani and $\epsilon$ Ophuichi. For non-radial modes in these stars, as identified from the asymptotic approximation, we obtain the azimuthal orders and a constraint on the inclination angle. The latter combined with the projected rotational velocity gives an estimate of the stellar rotational frequency.

The results of the line-profile analysis for $\beta$~Hydri, $\delta$ Eridani and $\epsilon$~Ophiuchi gave us confidence that we can perform a quantitative comparison between line-profile variations from observations and synthetic spectra, which provides the azimuthal order and for non-radial oscillators, the inclination angle and with that the surface rotational frequency. Therefore, we perform the same analysis for $\eta$~Serpentis (K0III), for which no previous mode identification was available (Section 5).

\section{Observations}
For $\epsilon$~Ophiuchi and $\eta$~Serpentis we have spectra from the fiber-fed \'{e}chelle spectrograph CORALIE mounted on the Swiss 1.2~m Euler telescope at La Silla (ESO, Chile) at our disposal. These were obtained during a two-site campaign using CORALIE and ELODIE, the fiber-fed spectrograph mounted on the French 1.93~m telescope at Observatoire de Haute Provence, France, during the summer of 2003. The ELODIE spectra are available, but their cross-correlation profiles have a resolution of 1 km\,s$^{-1}$ which is coarse compared to the 0.1 km\,s$^{-1}$ resolution of the CORALIE spectra, and are hence not used for the present analysis. The observations of $\delta$~Eridani were taken with CORALIE during a twelve day campaign in November 2001. 
For $\beta$~Hydri we use spectra obtained with the fiber-fed \'{e}chelle spectrograph HARPS mounted on the 3.6~m telescope at La Silla (ESO, Chili), which were obtained during a two-site campaign using HARPS and UCLES, the \'{e}chelle spectrograph mounted on the 3.9~m Anglo-Australian Telescope, Siding Spring Observatory, Australia, in September 2005. Spectra from UCLES are contaminated with iodine absorption lines, used to obtain accurate radial velocity variations, and are therefore not useful for the present analysis. For a detailed description of the data we refer to the publications describing the observations of solar-like oscillations in these stars, \citet{deridder2006,barban2004,carrier2002,carrier2003,bedding2007}, respectively. 
Stellar parameters of the four stars are listed in Table~\ref{propstar}.

We note here that we have data for $\xi$ Hydrae at our disposal, but as already shown by \citet{hekker2006} the signal-to-noise ratio of the oscillations is too low for the present analysis and therefore we did not include $\xi$ Hydrae in the current investigation.


\begin{figure*}
\begin{minipage}{4.2cm}
\centering
\includegraphics[width=4.2cm]{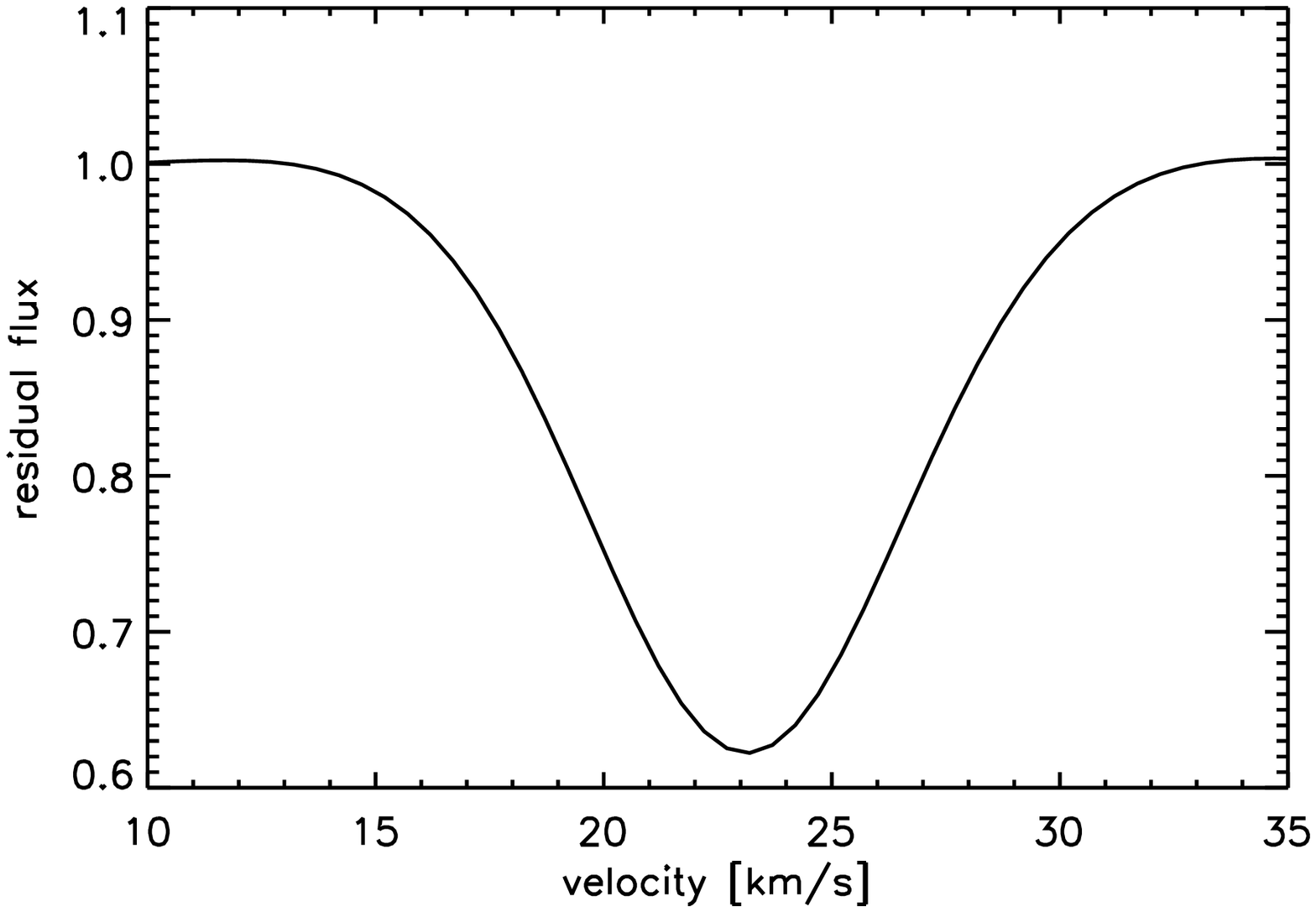}
\end{minipage}
\hfill
\begin{minipage}{4.2cm}
\centering
\includegraphics[width=4.2cm]{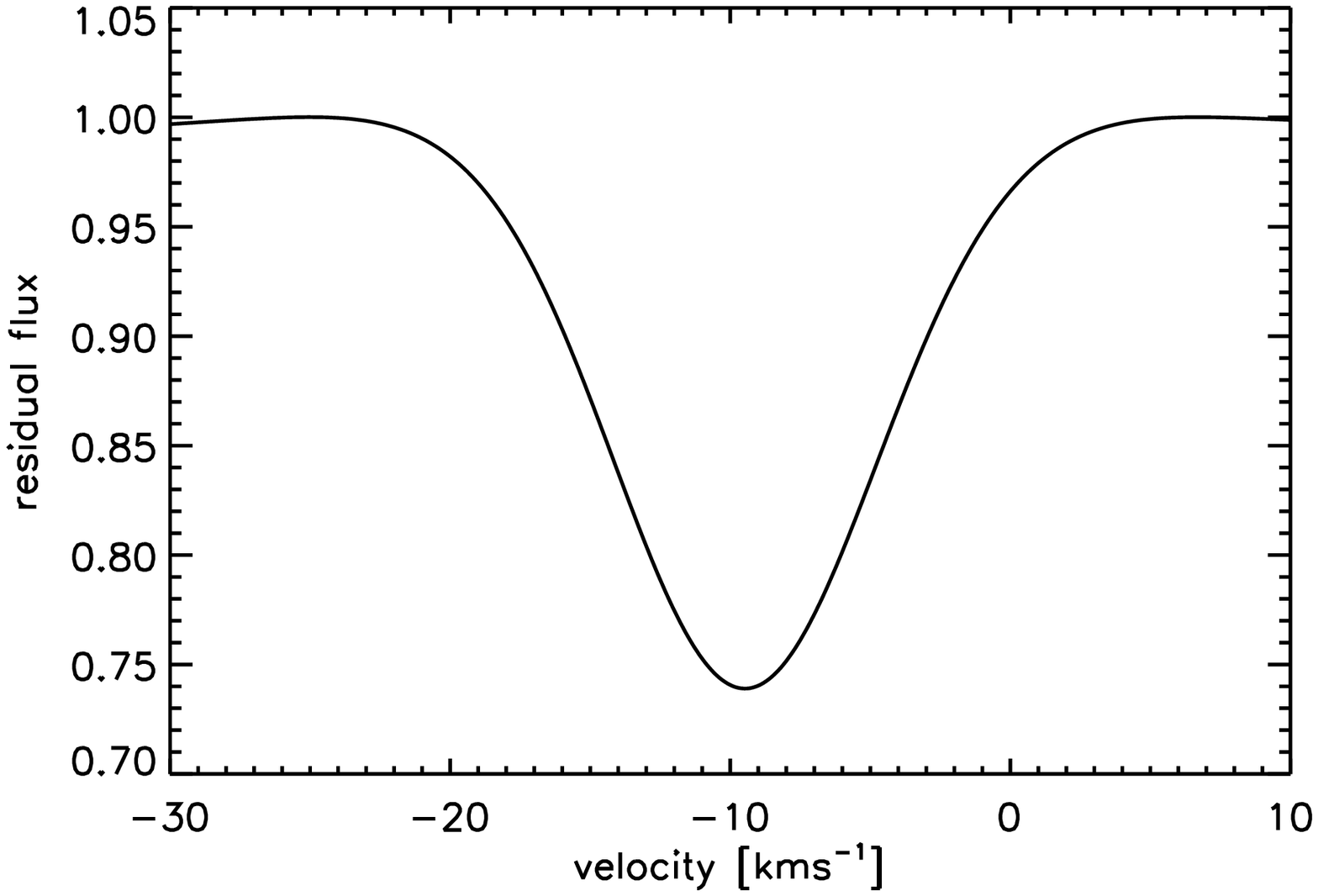}
\end{minipage}
\hfill
\begin{minipage}{4.2cm}
\centering
\includegraphics[width=4.2cm]{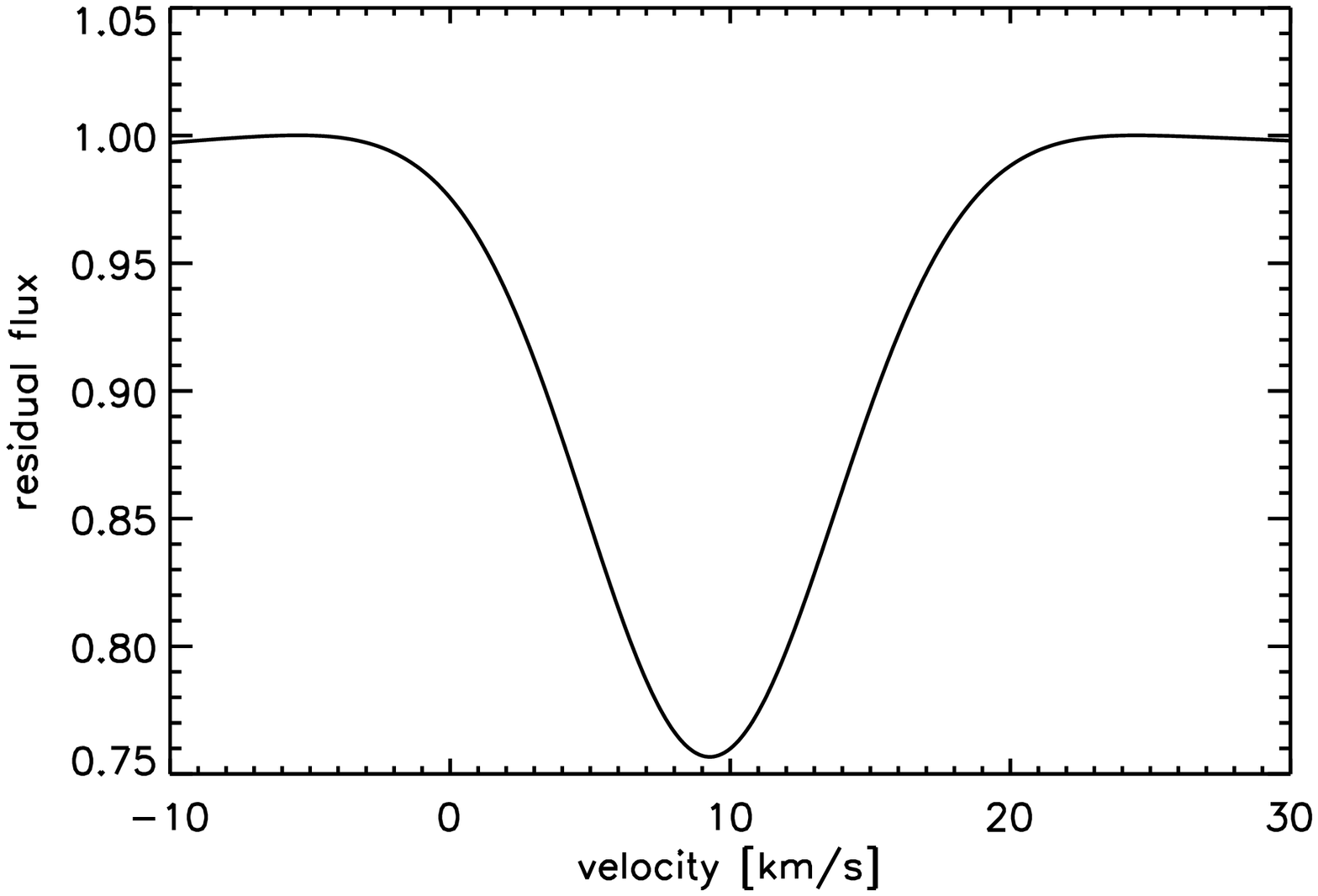}
\end{minipage}
\hfill
\begin{minipage}{4.2cm}
\centering
\includegraphics[width=4.2cm]{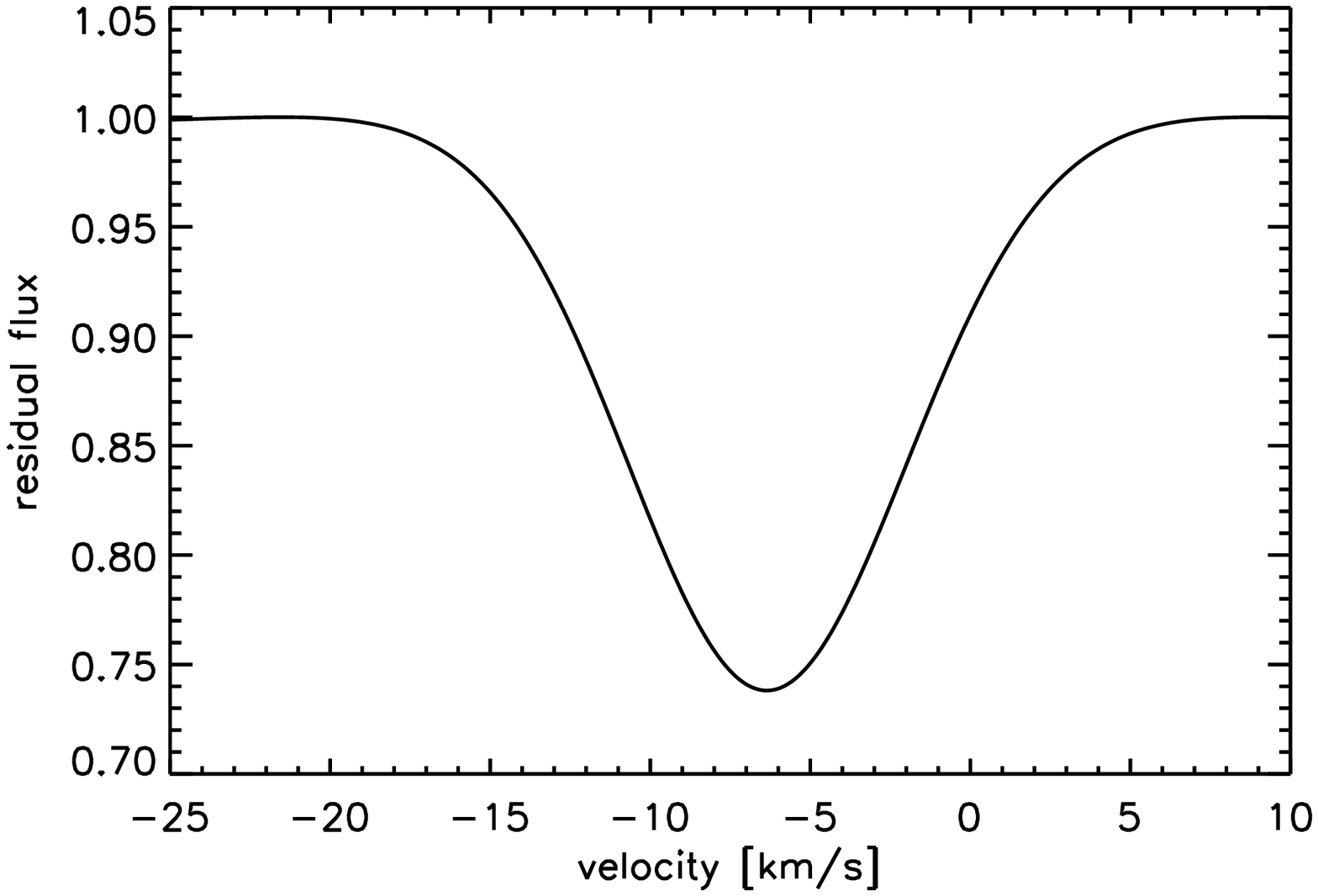}
\end{minipage}
\hfill
\begin{minipage}{4.2cm}
\centering
\includegraphics[width=4.2cm]{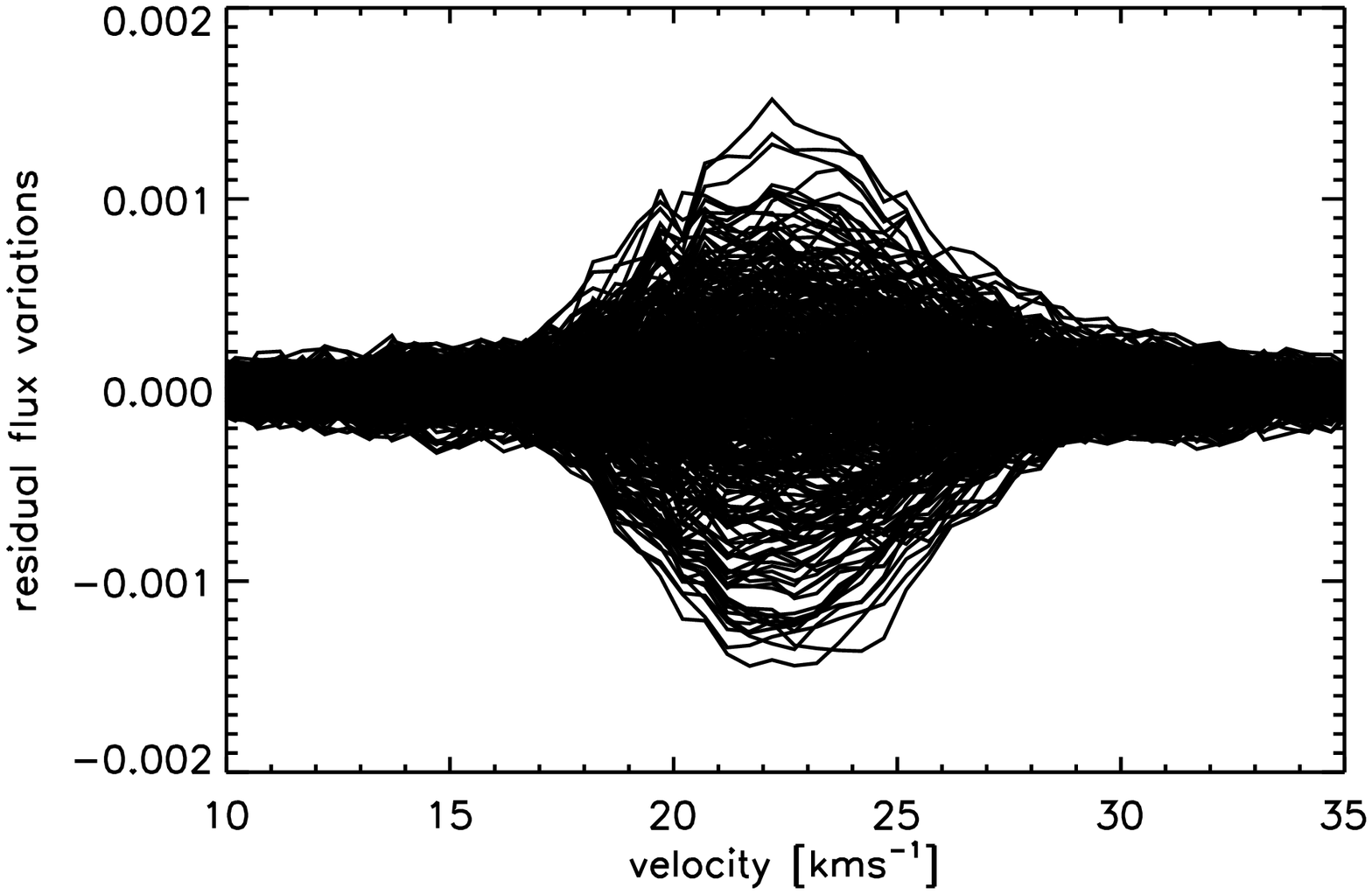}
\end{minipage}
\hfill
\begin{minipage}{4.2cm}
\centering
\includegraphics[width=4.2cm]{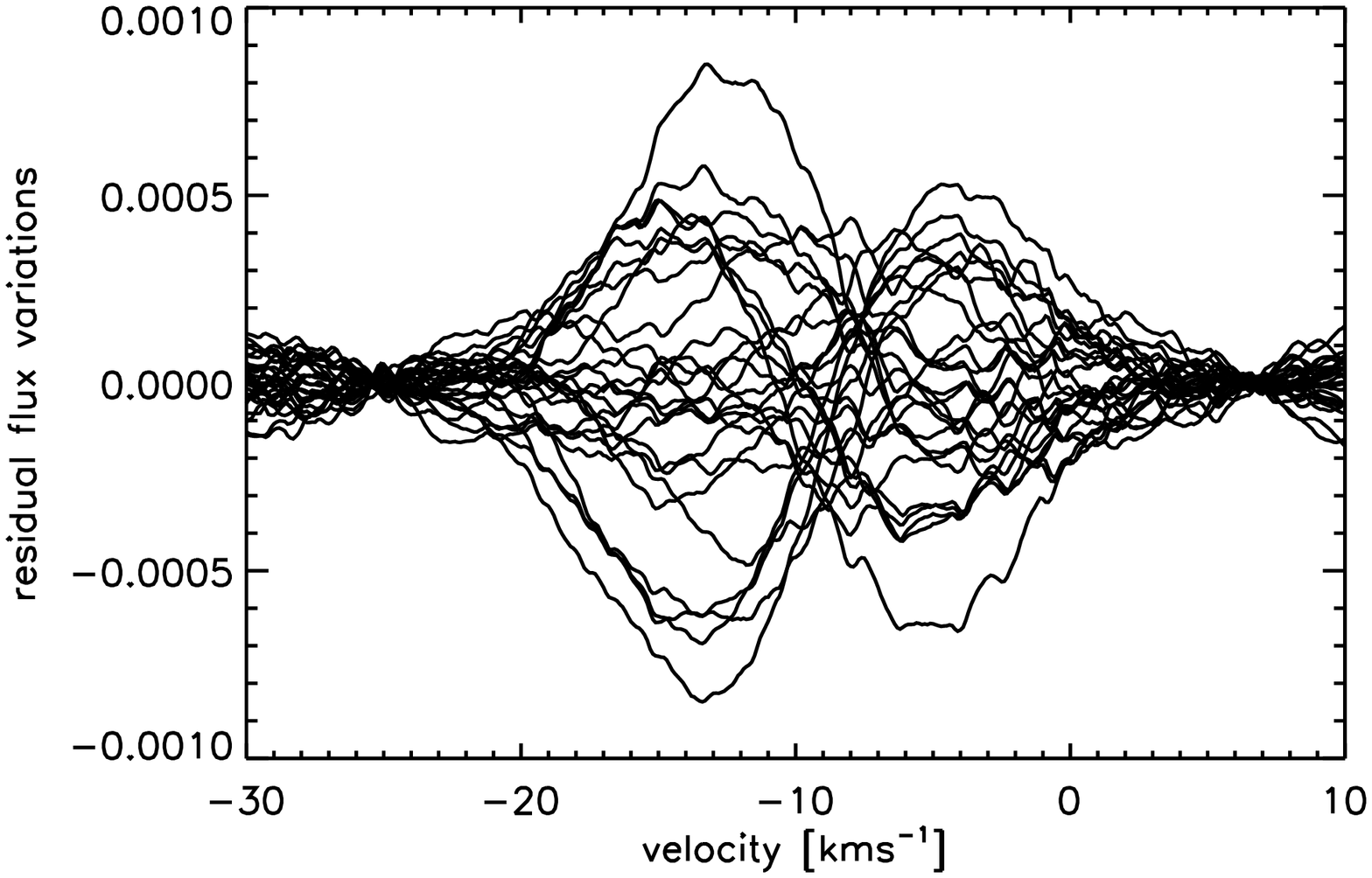}
\end{minipage}
\hfill
\begin{minipage}{4.2cm}
\centering
\includegraphics[width=4.2cm]{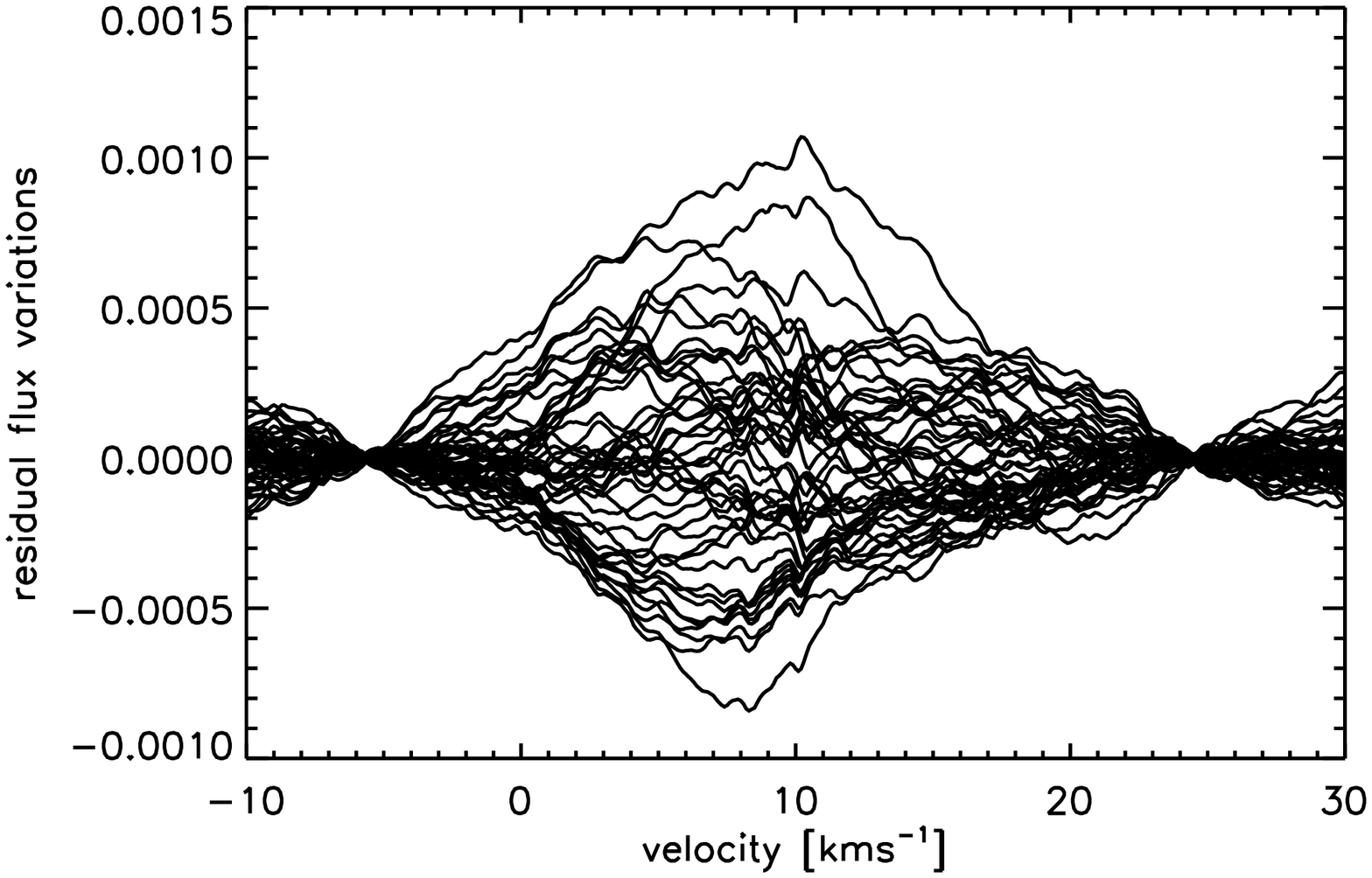}
\end{minipage}
\hfill
\begin{minipage}{4.2cm}
\centering
\includegraphics[width=4.2cm]{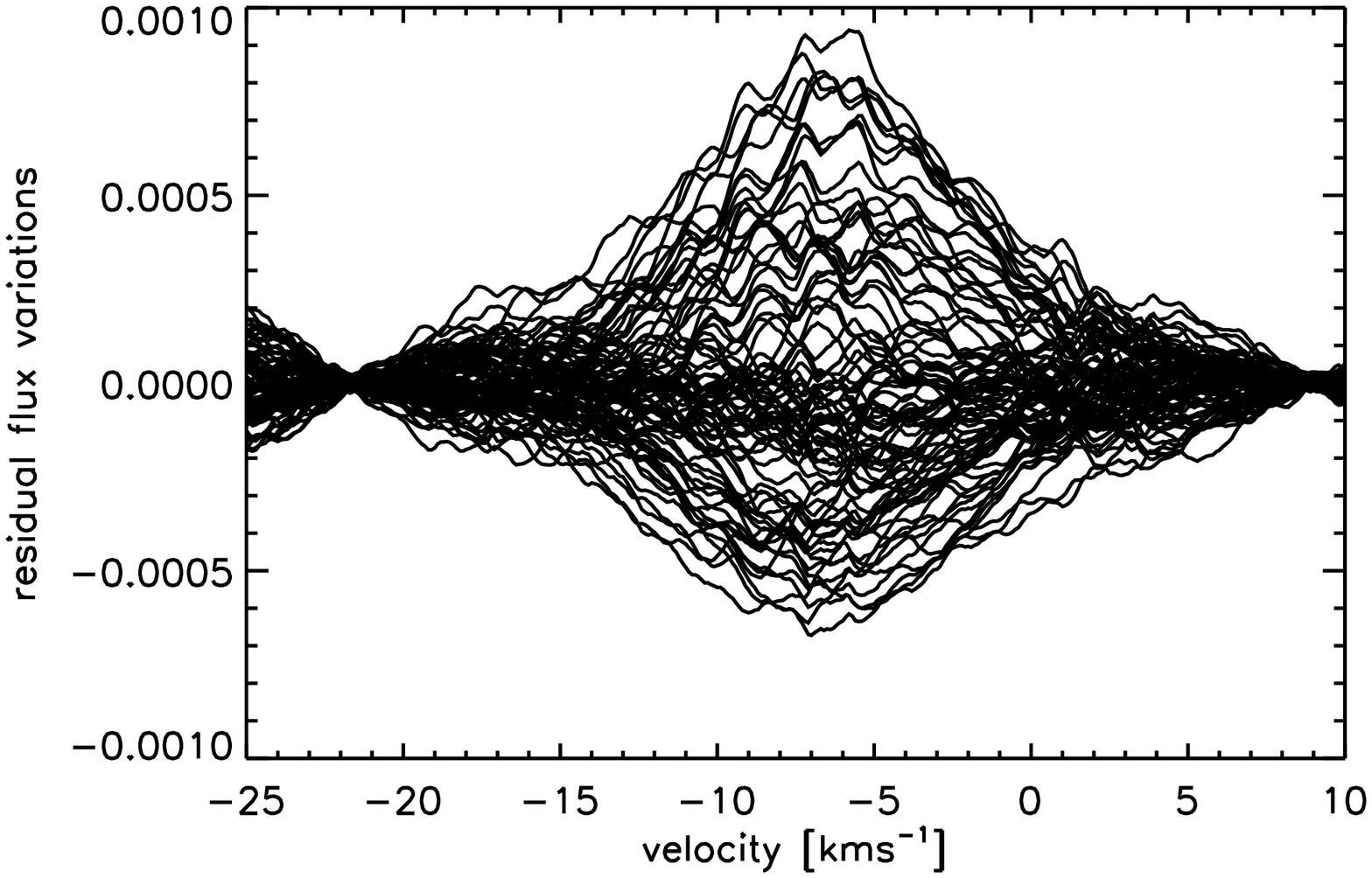}
\end{minipage}
\hfill
\begin{minipage}{4.2cm}
\centering
\includegraphics[width=4.2cm]{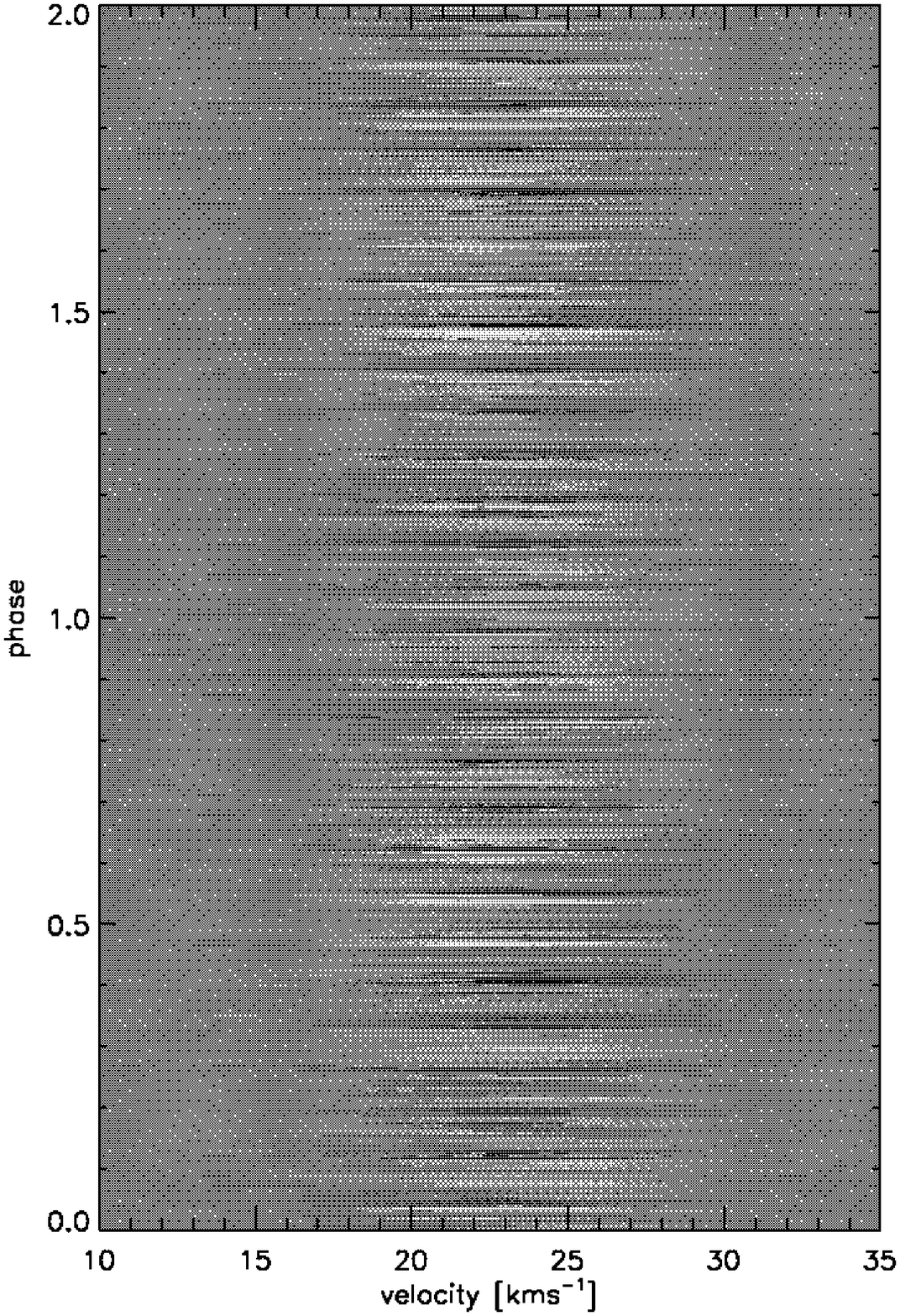}
\end{minipage}
\hfill
\begin{minipage}{4.2cm}
\centering
\includegraphics[width=4.2cm]{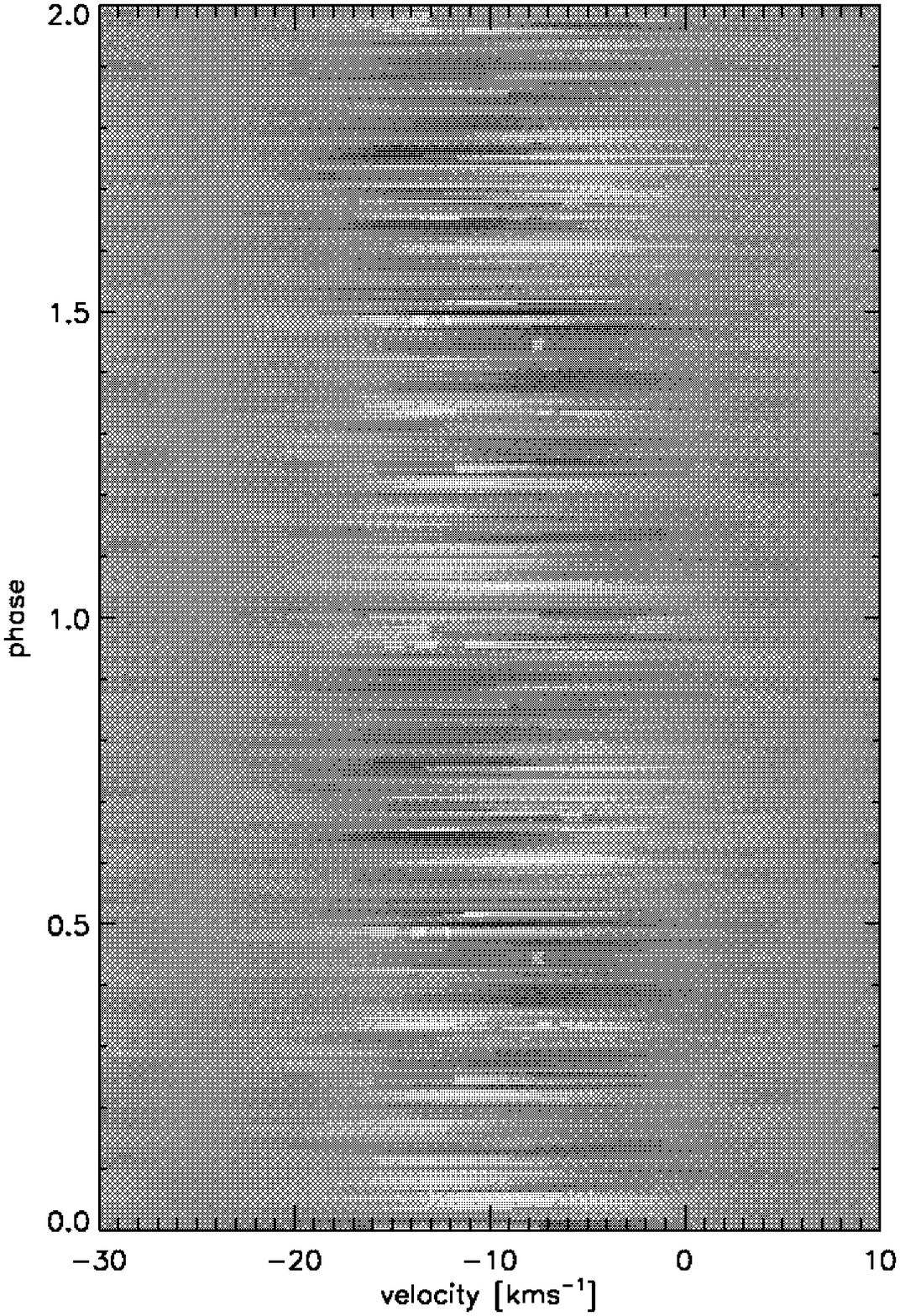}
\end{minipage}
\hfill
\begin{minipage}{4.2cm}
\centering
\includegraphics[width=4.2cm]{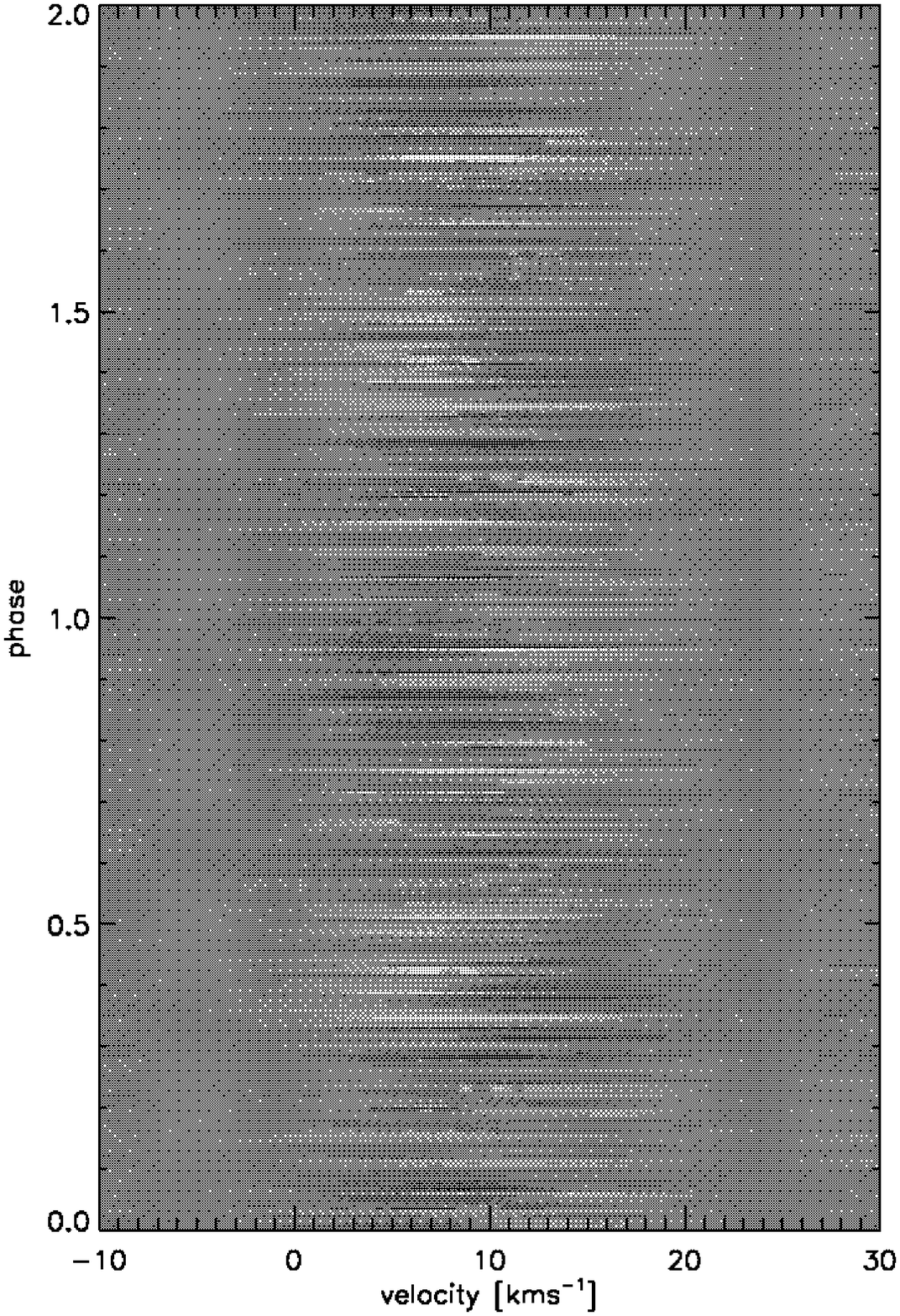}
\end{minipage}
\hfill
\begin{minipage}{4.2cm}
\centering
\includegraphics[width=4.2cm]{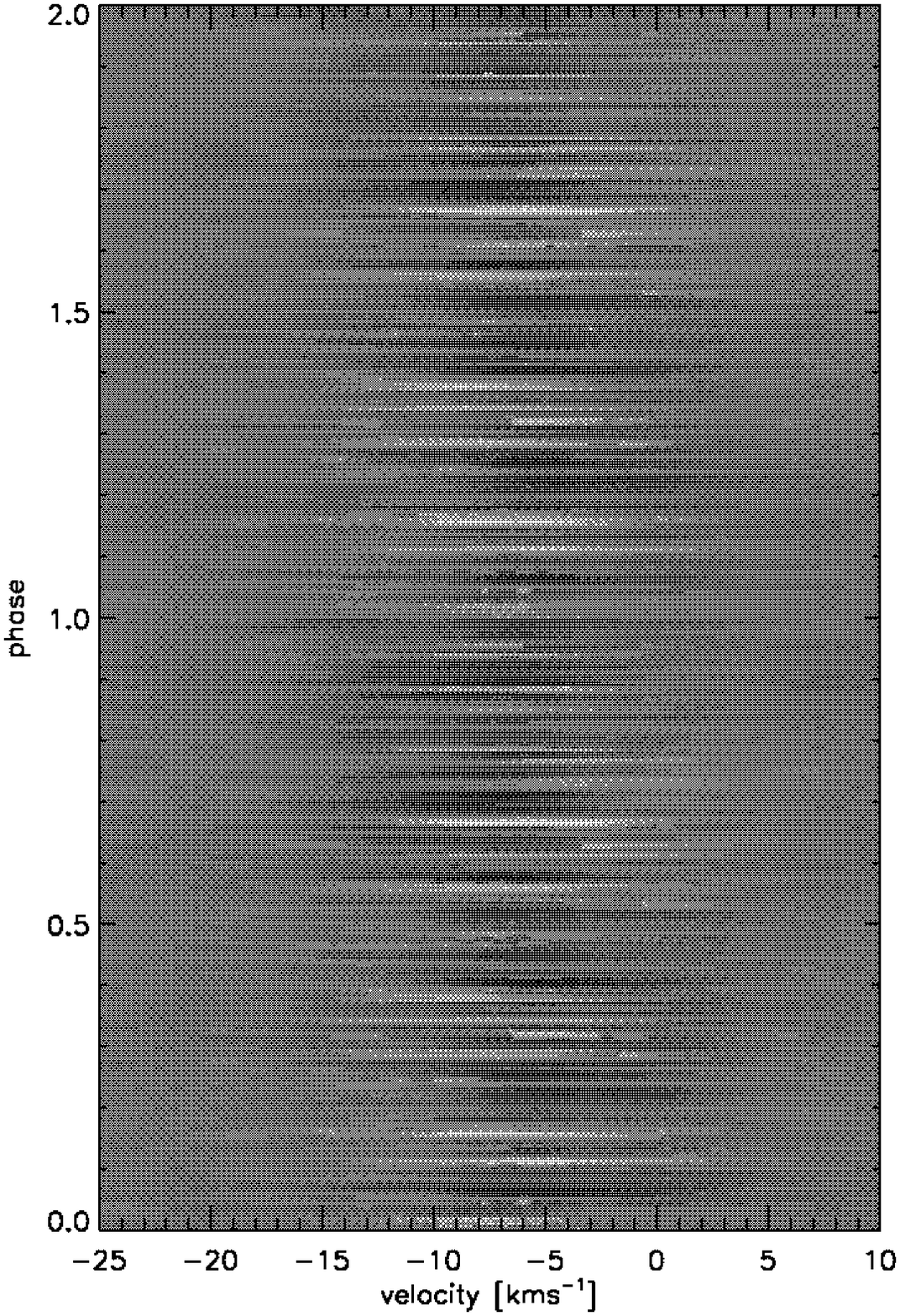}
\end{minipage}
\caption{Average cross correlation functions (top), isolated line-profile variations after subtraction of the average CCF (middle, only one night is displayed for clarity) and grayscale plots of the isolated line-profile variations as a function of phase of the respective dominant frequencies, where darker regions indicate negative values, while lighter regions indicate positive values (bottom). From left to right $\beta$~Hydri, $\epsilon$~Ophiuchi, $\eta$~Serpentis and $\delta$~Eridani. }
\label{CCFresgs}
\end{figure*}

\section{Improved analysis}
\citet{hekker2006} used the INTER-TACOS (INTERpreter for the Treatment, the Analysis and the COrrelation of Spectra) software package developed at Geneva Observatory \citep{baranne1996} to compute cross-correlation functions (CCFs) of the observed spectra. The CCFs are computed using a mask containing absorption lines of a K0 star for $\epsilon$~Ophiuchi, $\eta$~Serpentis and $\delta$~Eridani and have a resolution of 0.1 km\,s$^{-1}$. For the HARPS data of $\beta$~Hydri we use the CCFs computed with the HARPS pipeline \citep{rupprecht2004}, which uses a mask with absorption lines of a G2 star. These CCFs have a resolution of 0.5 km\,s$^{-1}$.

Here we use these CCFs as a starting point of a line-profile analysis, but we apply three corrections before attempting to analyse the line-profile variations. These corrections appeared to be necessary, as effects not intrinsic to the star (instrumental effects / changing weather conditions) cause additional changes in the line profiles on a night by night basis. These effects hampered a quantitative analysis by \citet{hekker2006}. The following corrections are applied to isolate the line-profile variations due to oscillations from the extrinsic ones:
\begin{itemize}
\item The first correction concerns a continuum normalisation. Continuum on both sides of each CCF is selected and a linear polynomial is fitted through these continuum points. The full cross-correlation profile is then divided by this linear fit. \newline
\item The second correction involves the isolation of the line-profile variations and removal of outliers. For each night of observations we calculate for each velocity bin the average CCF value and its standard deviation. In each bin the average CCF is subtracted from the observed continuum corrected CCF to isolate the variations. Outliers are identified as values larger than three times the standard deviation. CCFs with outliers in one or more velocity bins are fully discarded. \newline
\item Finally, the flux variations in each velocity bin are imposed on the mean CCF of all observations. This is not strictly necessary to perform line-profile analysis, but we aim to use moments to obtain the oscillation frequencies and therefore we need to have a line profile rather than the residual profile.
\end{itemize}

The resulting average CCFs and isolated line-profile variations for all four stars are shown in Fig.~\ref{CCFresgs} together with grayscale plots of the isolated line-profile variations.

Extensive tests have been performed to investigate the influence of these corrections and removal of outliers, in particular to confirm that these indeed remove effects not intrinsic to the star and that these do not influence the intrinsic behaviour. These tests indicate that the results presented here are robust to degrading the signal and to changes in the average profiles over night due to for instance changed weather conditions or instrumental effects.

With FAMIAS \citep{zima2008}, we compute frequencies from the first moment \citep{aerts1992} for the new CCFs, similar to what was done by \citet{hekker2006} for the uncorrected CCFs. In general we recover the same frequencies (or 1 day aliases) as \citet{hekker2006} for $\epsilon$ Ophiuchi, $\eta$ Serpentis and as \citet{bedding2007,carrier2002,carrier2003} for $\beta$ Hydri and $\delta$ Eridani, respectively. 

\section{Line-profile analysis}

The line-profile analysis is performed with the Fourier Parameter Fit (FPF) method \citep{zima2006}. The FPF method relies on a fit of the observational Fourier parameters across the line profile. For every detected pulsation frequency, the zero point, amplitude, and phase are computed for every velocity bin across the profile and compared with the same quantities from synthetic profiles. This procedure only relies on the velocity eigenfunction at the stellar surface and is therefore model independent (see Chapter 6 of \citet{aerts2009}). We use the standard sign convention for the azimuthal order $m$ in FAMIAS, i.e., $m$ $>$ 0 implies pro-grade modes.

One has to keep in mind that this method is mainly sensitive to the azimuthal order ($m$) of the mode and much less to its harmonic degree ($\ell$), as modes with the same order but similar degree will give essentially the same amplitude and phase distributions \citep{zima2006}. 

For the present analysis we use the frequencies of the first moment of the new CCFs and stellar parameters as listed in Table~\ref{propstar} as input parameters. 
We first computed pulsationally independent parameters, i.e., equivalent width (EW), macroturbulent broadening ($\xi_{\rm macro}$), projected rotational velocity ($\upsilon \sin i$) and radial velocity (RV) and fix these. The inclination angle is a free parameter in all calculations. Then we use the FPF method to fit the amplitude and phase distributions either for modes for which we know the degree from previous determinations or for a grid of modes with degree $0 \leq \ell \leq 4$, and order $- \ell \leq m \leq \ell$. Modes up to degree 4 were investigated as in spectroscopy the partial cancellation effect would theoretically allow this. For examples of amplitude and phase distributions, see e.g., Fig.~\ref{ZAPbhydri1}. 

Although we investigate stochastic oscillations, we did not include damping and excitation in the synthetic profiles, but take the stochastic effects into account in the interpretation of the results. The main reason for this is that we know that damping and excitation can cause asymmetries in the amplitude and phase distribution \citep{hekker2006}, but the exact nature of these asymmetries varies per realisation. In their Fig.~10, \citet{hekker2006} show examples of amplitude and phase distributions of simulated noise free line profiles with a two day damping time. For each of the modes $\ell$ = 0, 1, 2 and $m$ = 0 - $\ell$ they show 10 realisations. From these simulations it is clear that for different realisations of the stochastically damped and re-excited oscillations, the height of the amplitude distributions varies for all modes. In addition to that, the centroids of the distributions for modes with $m$ $\neq$ 0 can be shifted from the laboratory wavelength. The asymmetries in the amplitude profiles are most pronounced for sectorial modes with $\ell$ = $|m|$, but are also present for other $m$ $\neq$ 0 modes. Moreover, from simulations of spectra with signal-to-noise ratios of 50, 100 and 150, we concluded that the asymmetries in the amplitude distribution are also present for modes with $m$ = 0.
In the interpretation of our results we take the asymmetries and shifts into account by shifting the computed amplitude distributions such that their central minima overlap with the central minimum of the observed distribution. The same shift is applied to the computed phase distributions.

To identify the best fit to the data, $\chi^2$ values have been computed, as foreseen in FAMIAS. These are used to select the fits with $\chi^2$ below 1. However, it turned out that a reasonably large number of  profiles satisfies this requirement and that among these best fits the optimum $m$ value can differ when investigating only the central part of the line, or when the full line profile is taken into account. Therefore, we discuss the results of our line-profile modelling from visual inspection of the amplitude and phase distributions across the line profiles as it has been done previously in the literature, see e.g. \citet{telting1997} and \citet{briquet2005,briquet2009} for $\beta$ Cephei, $\theta$~Ophiuchi and HD180642, respectively.

\section{Results}

The analysis of line-profile variations for solar-like oscillations in evolved stars as discussed above is applied first on stars for which mode identification has already been performed, i.e., $\beta$~Hydri, $\delta$ Eridani and $\epsilon$~Ophiuchi. 

\subsection{$\beta$~Hydri}

\begin{table}
\begin{minipage}{\linewidth}
\caption{Global line profile parameters as computed with FAMIAS and used for the line profile analysis. Terminology similar to the one used in Table~\ref{propstar}.}
\label{inppar}
\centering
\begin{tabular}{lcccc}
\hline\hline
parameter & $\epsilon$ Ophiuchi & $\eta$ Serpentis & $\delta$ Eridani & $\beta$ Hydri\\
\hline
EW [km\,s$^{-1}$] & 3.0 & 2.7 & 2.8 & 3.2 \\
$\upsilon \sin i$ [km\,s$^{-1}$] & 5.7 & 5.6 & 4.9 & 4.3 \\
$\xi_{macro}$ [km\,s$^{-1}$] & 3.7 & 3.4 & 3.6 & 2.6 \\
\hline
\end{tabular}
\end{minipage}
\end{table}

\begin{figure*}
\begin{minipage}{5.6cm}
\centering
\includegraphics[width=5.5cm]{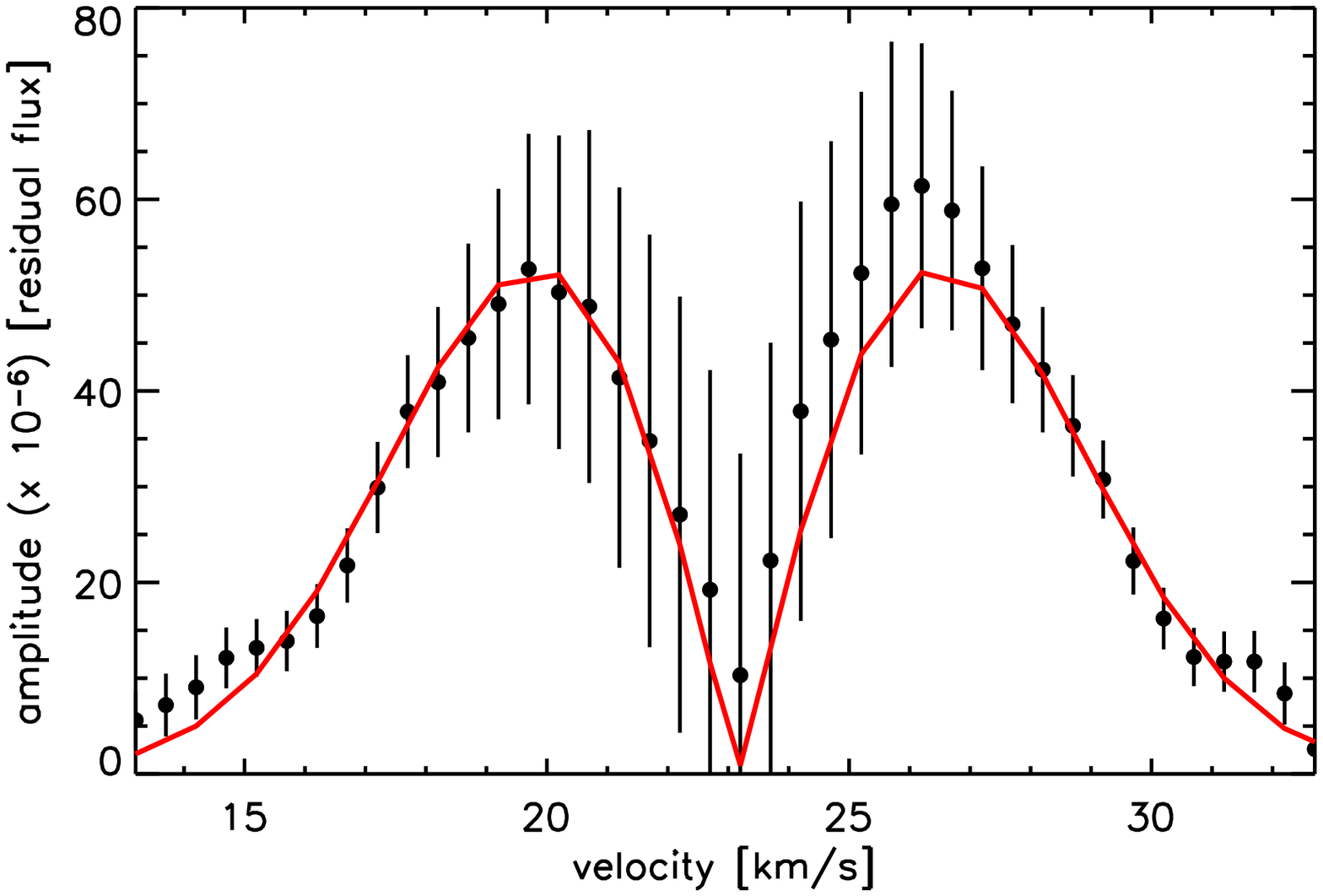}
\end{minipage}
\hfill
\begin{minipage}{5.6cm}
\centering
\includegraphics[width=5.5cm]{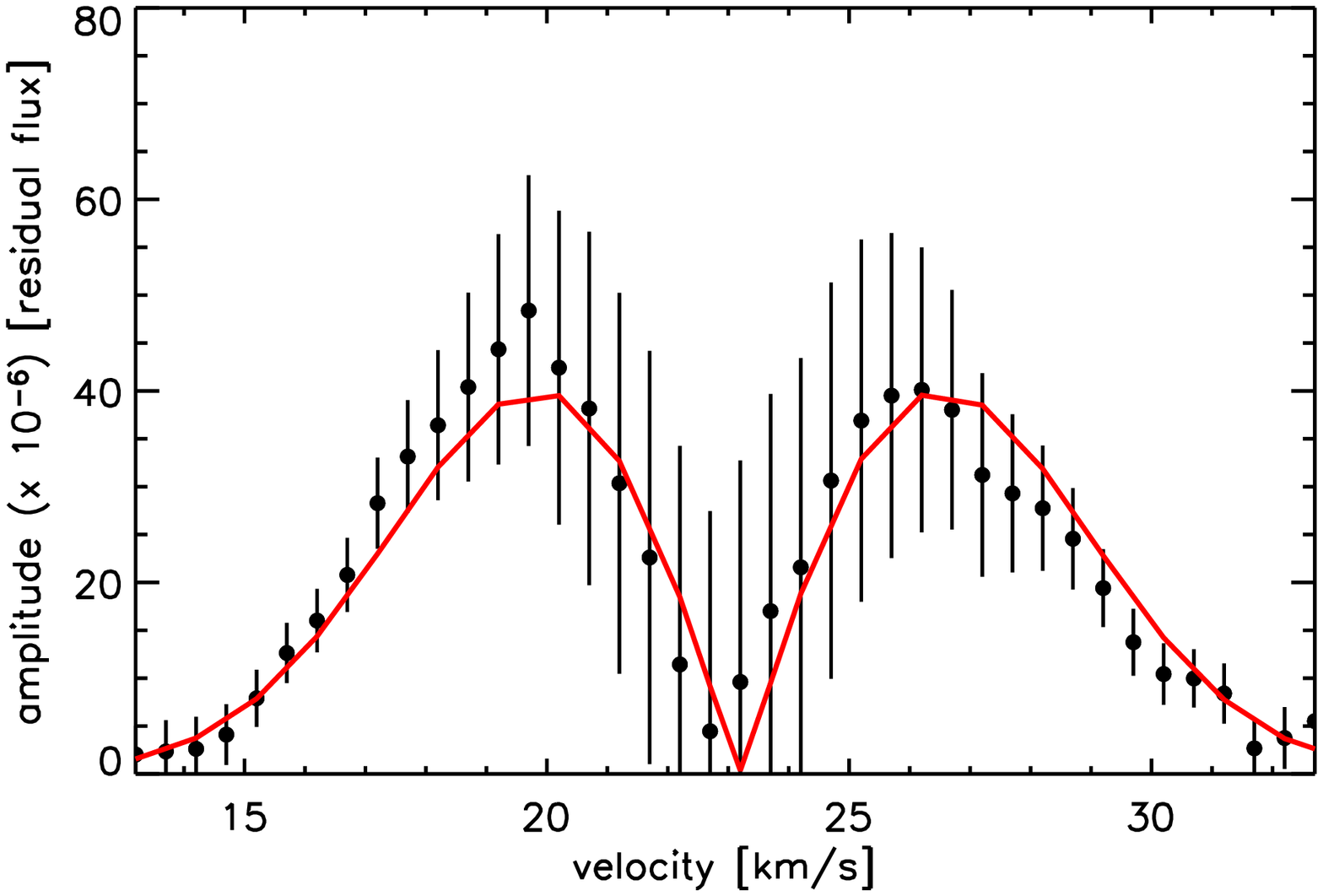}
\end{minipage}
\hfill
\begin{minipage}{5.6cm}
\centering
\includegraphics[width=5.5cm]{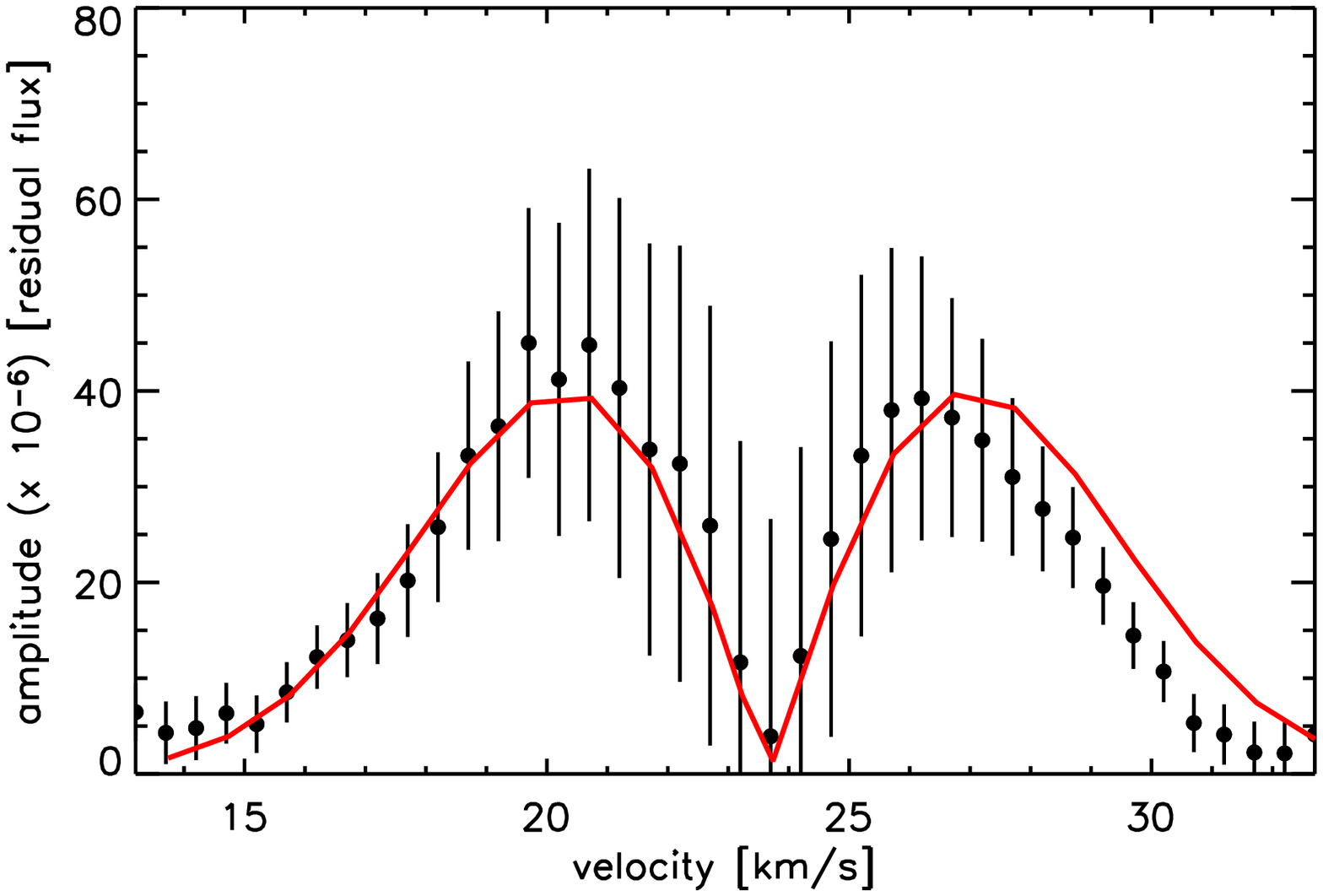}
\end{minipage}
\hfill
\begin{minipage}{5.6cm}
\centering
\includegraphics[width=5.5cm]{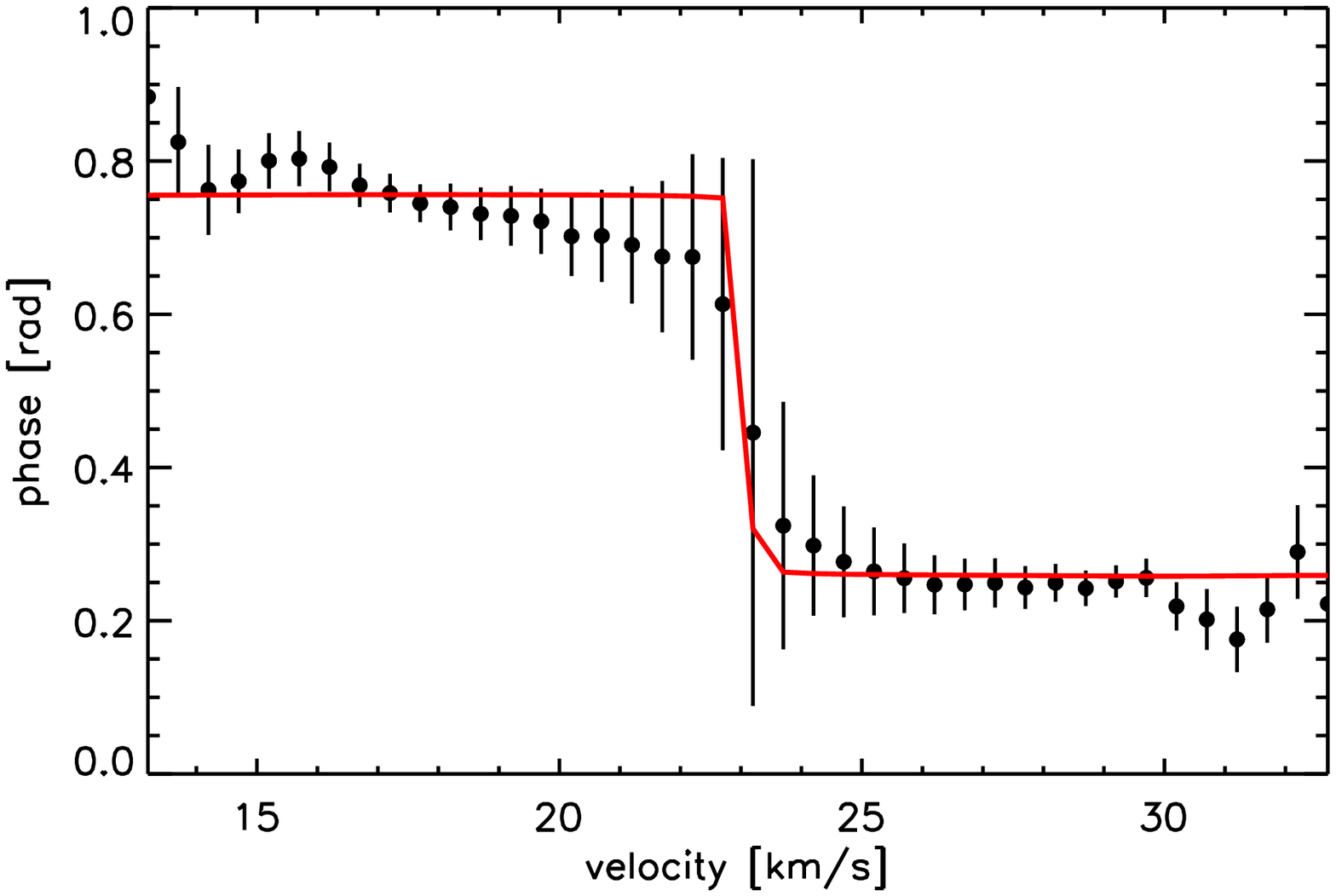}
\end{minipage}
\hfill
\begin{minipage}{5.6cm}
\centering
\includegraphics[width=5.5cm]{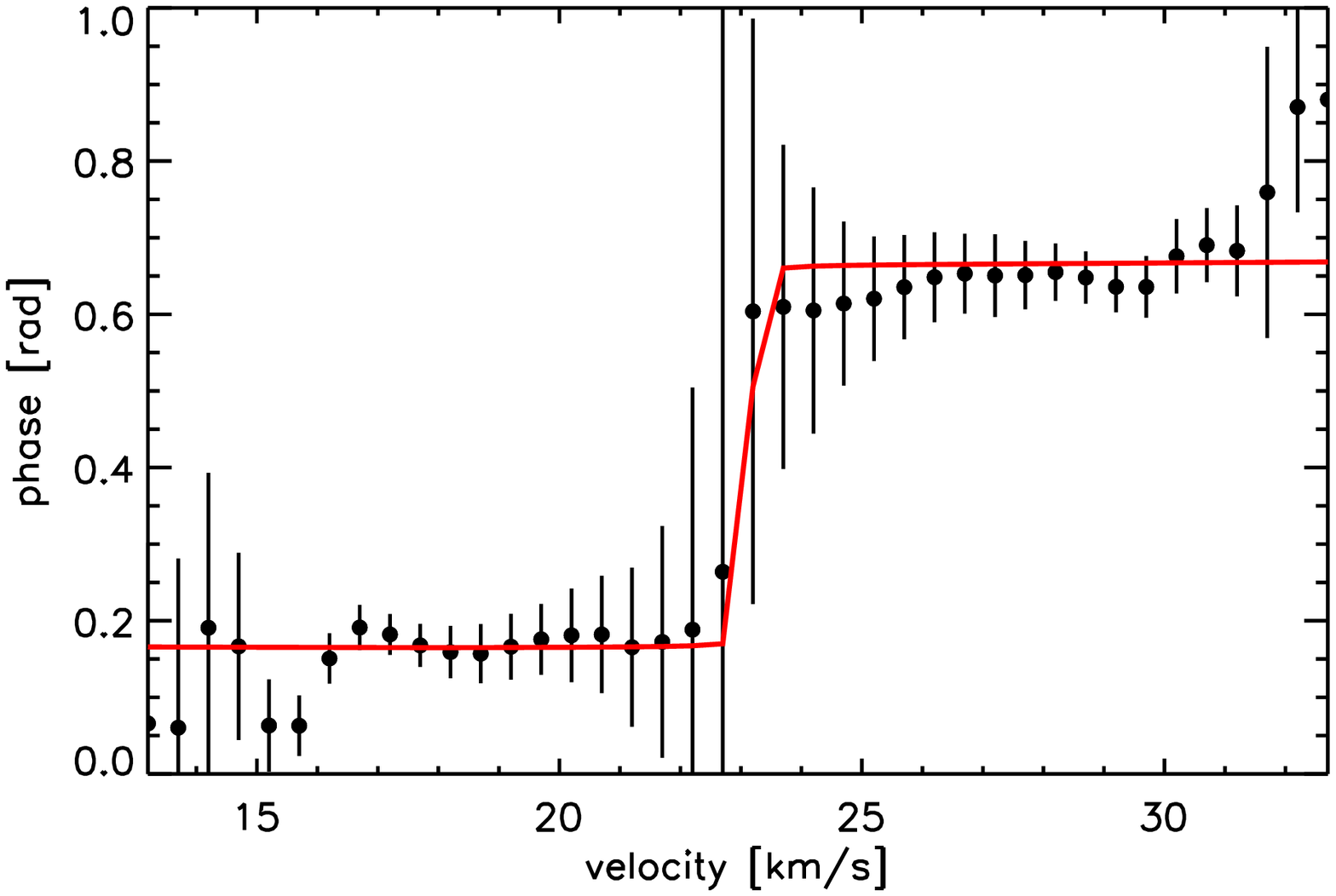}
\end{minipage}
\hfill
\begin{minipage}{5.6cm}
\centering
\includegraphics[width=5.5cm]{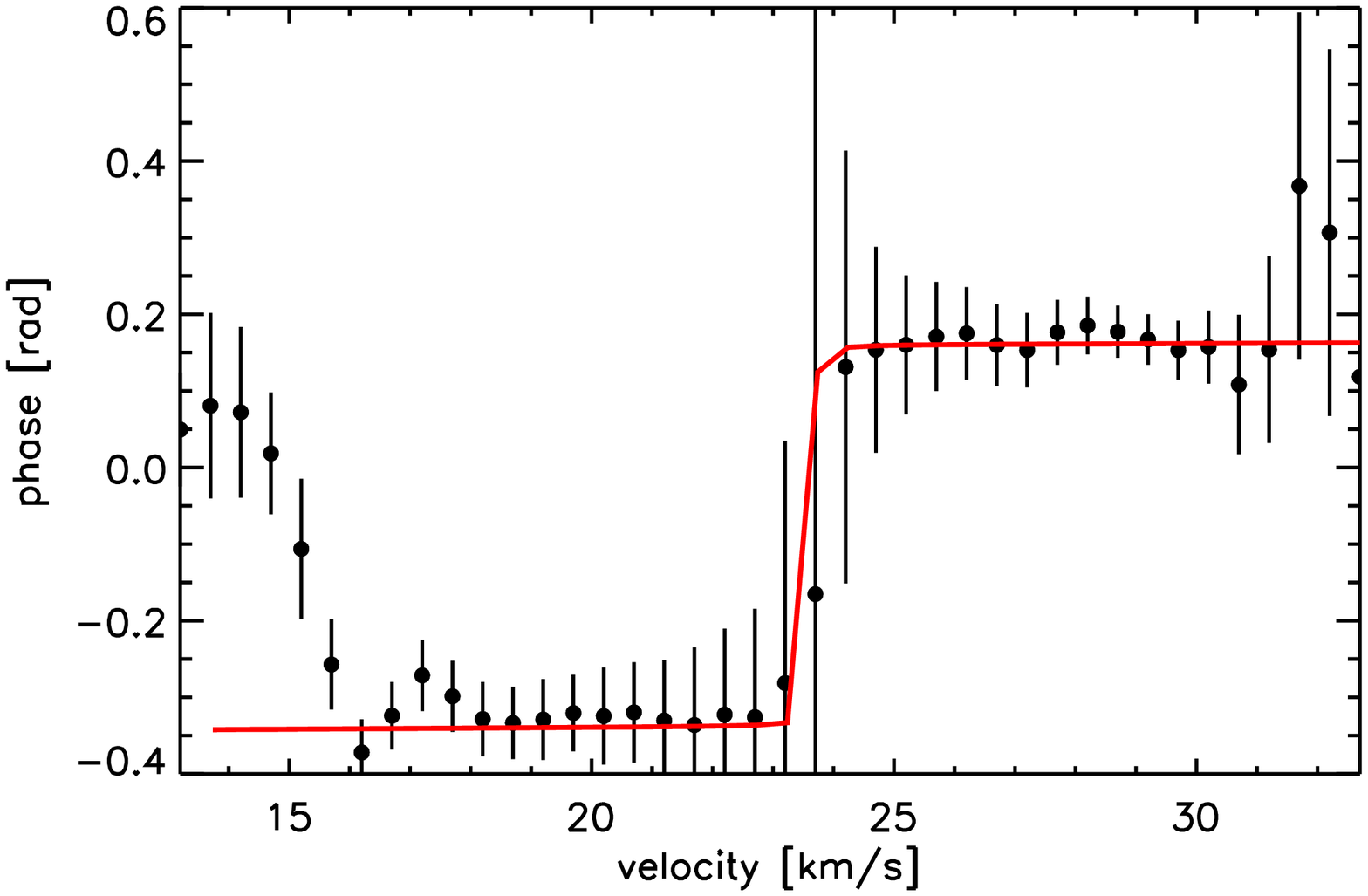}
\end{minipage}
\caption{Amplitude (top) and phase (bottom) distributions as a function of velocity across the line profile of $\beta$~Hydri for the three radial modes with frequencies obtained from the first moment $\langle \mathrm{v} \rangle$: $\nu_{\langle \mathrm{v} \rangle}=86.8$~c\,d$^{-1}$ (1004.4~$\mu$Hz) (left), $\nu_{\langle \mathrm{v} \rangle}=73.0$~c\,d$^{-1}$ (845.0~$\mu$Hz) (centre) and $\nu_{\langle \mathrm{v} \rangle}=96.7$~c\,d$^{-1}$ (1118.9~$\mu$Hz) (right). The mean radial velocity of the star is approximately $23.2$~km\,s$^{-1}$. The data is indicated with black dots and the $(\ell,m)$ = (0,0) fit is indicated in red solid line (colours only in the online version). In cases where the observed distributions are not centred around the midpoint of the CCF, the fits are shifted (see Section 4 for further explanation).}
\label{ZAPbhydri1}
\end{figure*}

\begin{figure*}
\begin{minipage}{5.6cm}
\centering
\includegraphics[width=5.5cm]{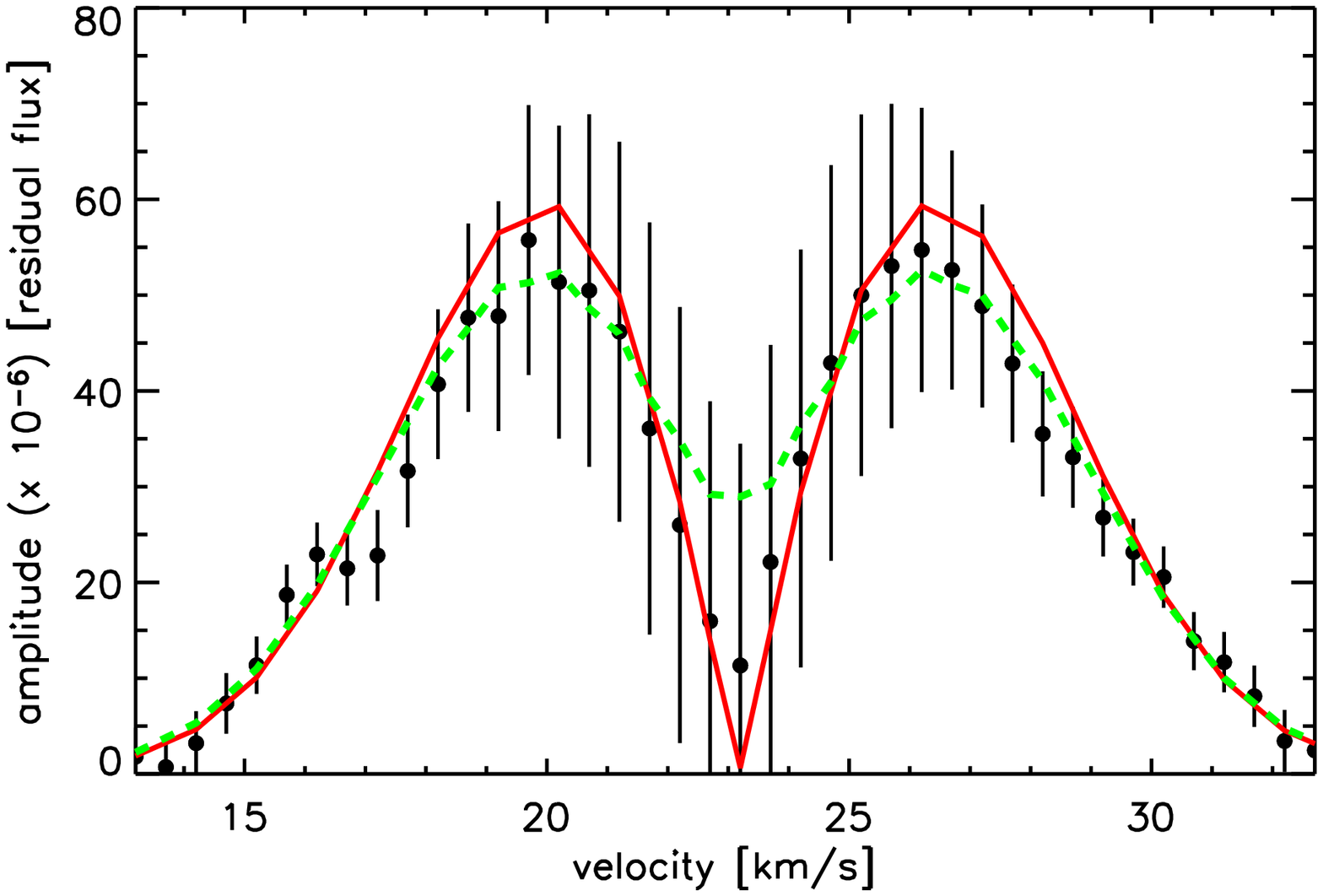}
\end{minipage}
\hfill
\begin{minipage}{5.6cm}
\centering
\includegraphics[width=5.5cm]{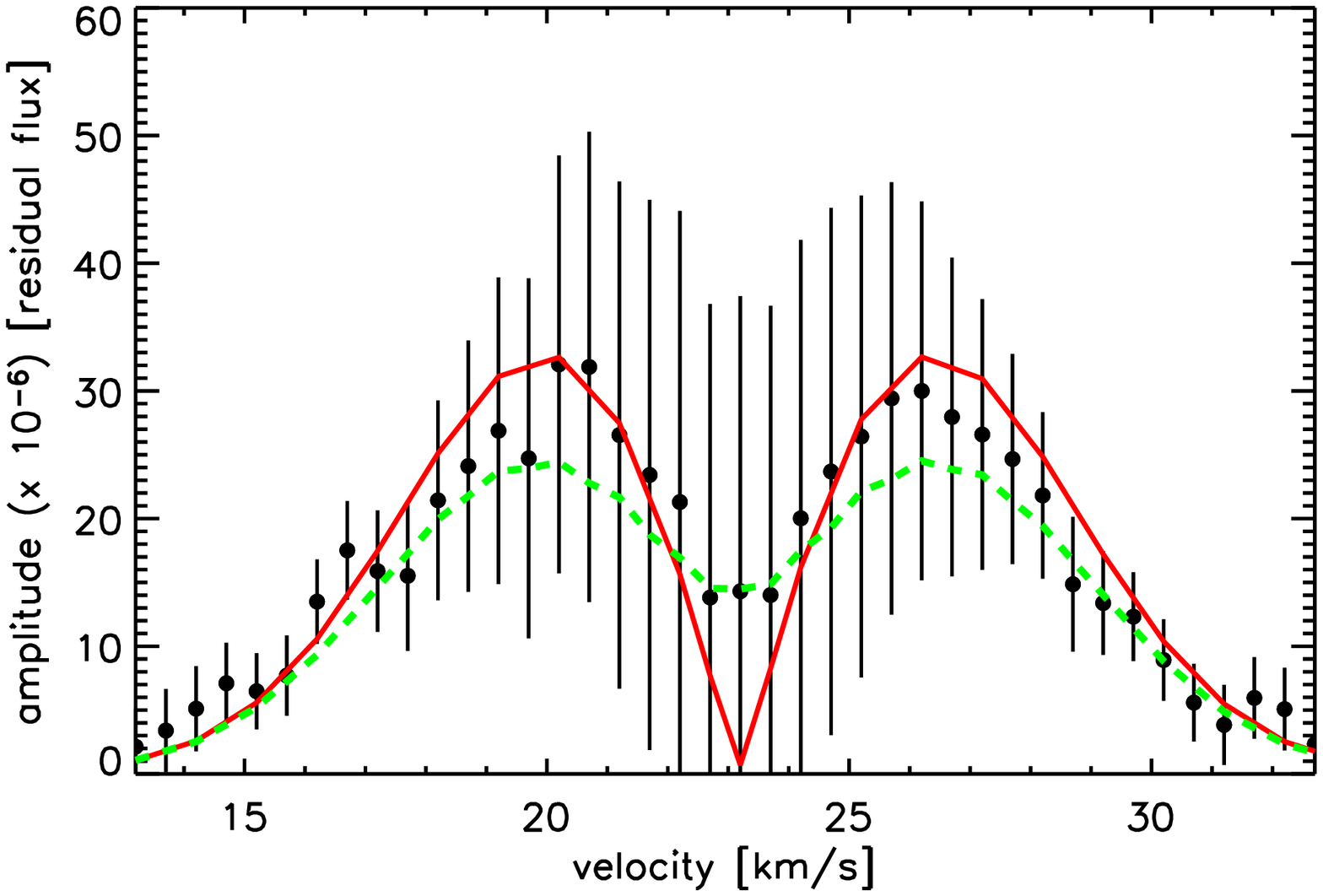}
\end{minipage}
\hfill
\begin{minipage}{5.6cm}
\centering
\includegraphics[width=5.5cm]{empty.eps}
\end{minipage}
\hfill
\begin{minipage}{5.6cm}
\centering
\includegraphics[width=5.5cm]{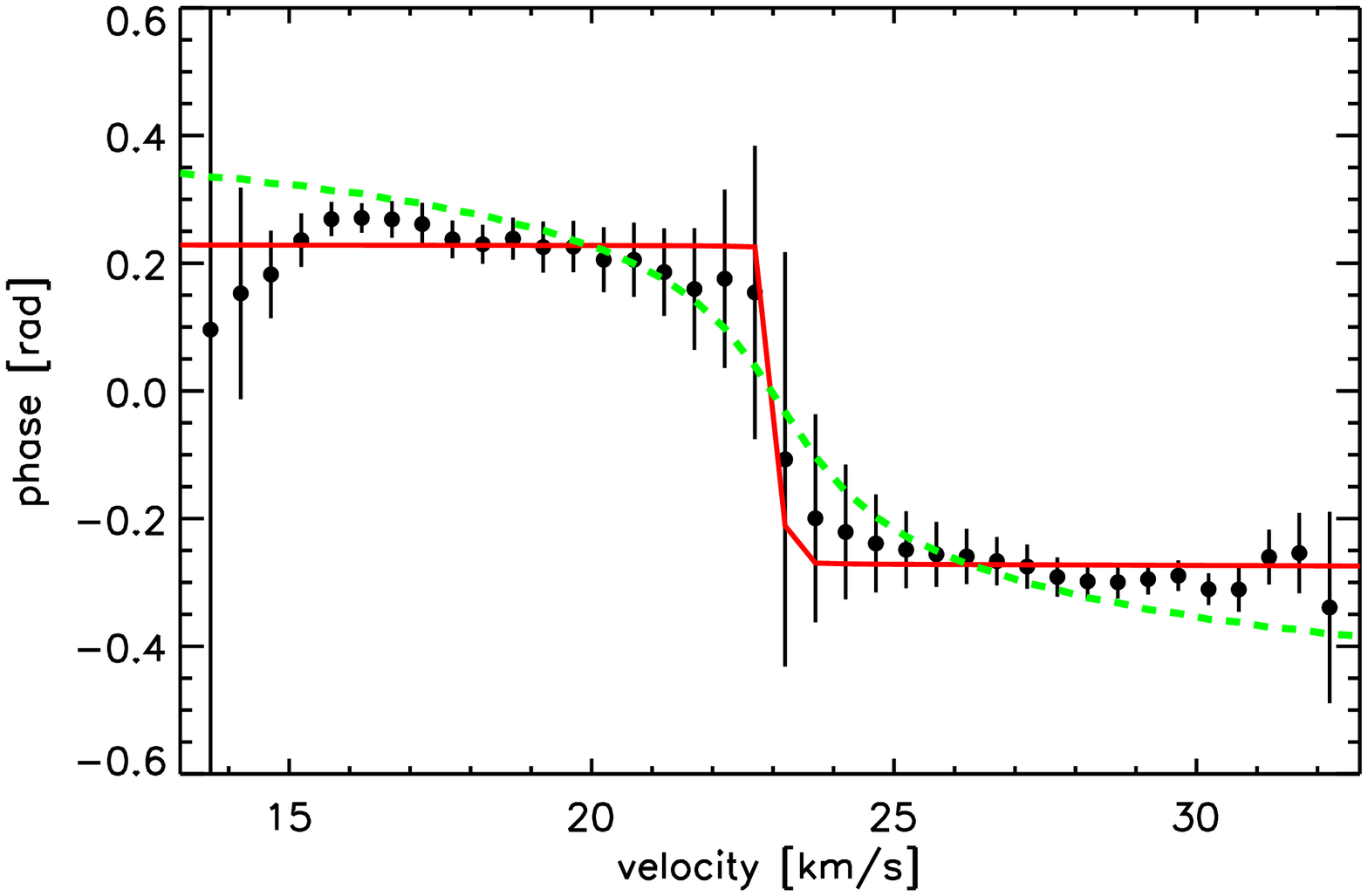}
\end{minipage}
\hfill
\begin{minipage}{5.6cm}
\centering
\includegraphics[width=5.5cm]{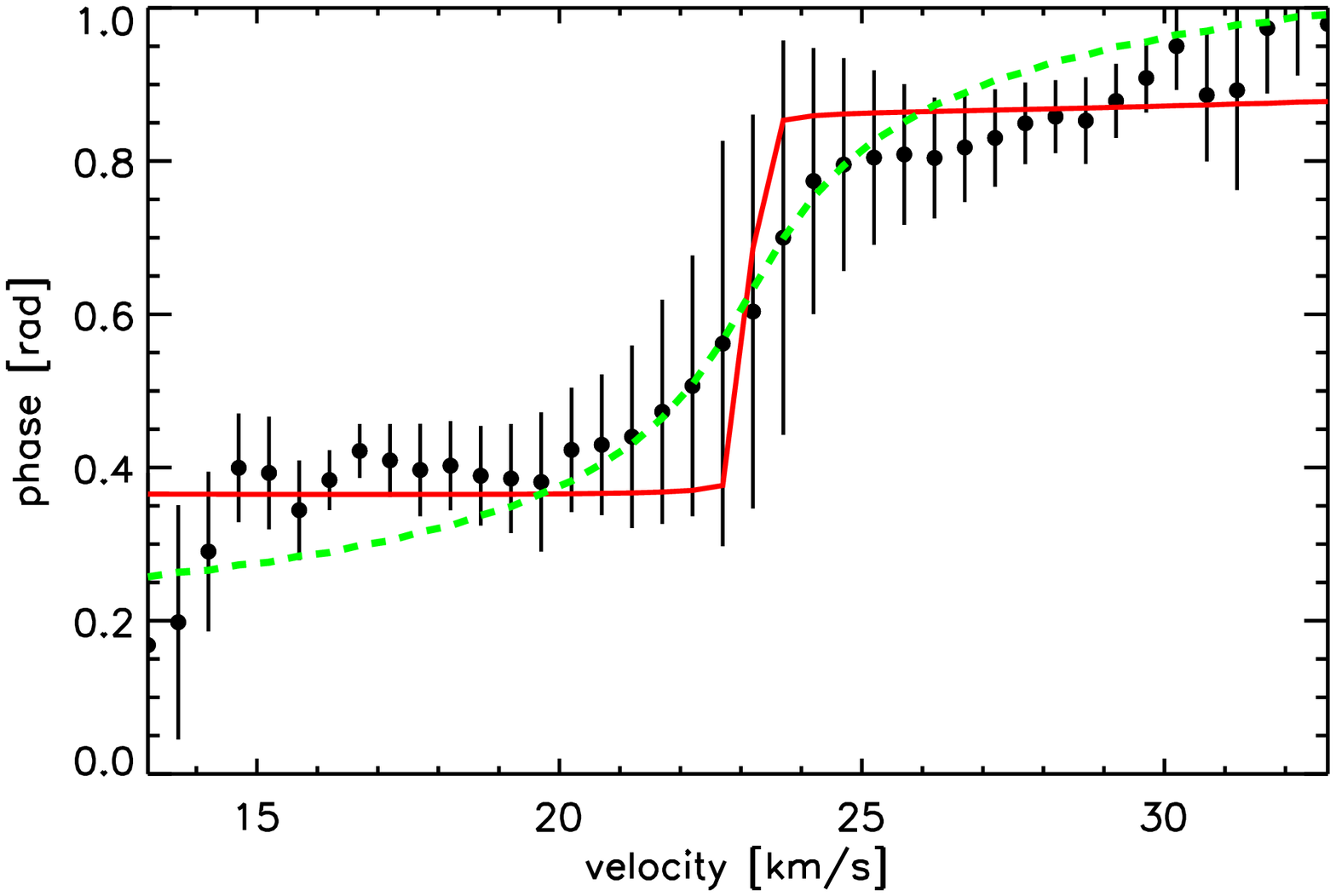}
\end{minipage}
\hfill
\begin{minipage}{5.6cm}
\centering
\includegraphics[width=5.5cm]{empty.eps}
\end{minipage}
\caption{Same as Fig.~\ref{ZAPbhydri1} for two frequencies obtained for $\beta$~Hydri from $\langle \mathrm{v} \rangle$: $\nu_{\langle \mathrm{v} \rangle}=89.3$~c\,d$^{-1}$ (1033.1~$\mu$Hz) (left), $\nu_{\langle \mathrm{v} \rangle}=103.0$~c\,d$^{-1}$ (1192.3~$\mu$Hz) (centre), for which the harmonic degrees have been determined previously to be $\ell$ = 1. In these panels the $(\ell,m)$ = (1,0) and $(\ell,m)$ = (1, $\pm$1) fits are indicated with red solid and green dashed lines, respectively. }
\label{ZAPbhydri2}
\end{figure*}

\begin{figure}
\begin{minipage}{\linewidth}
\centering
\includegraphics[width=\linewidth]{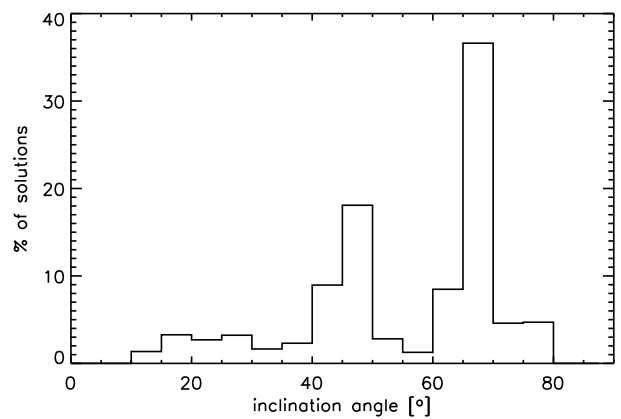}
\end{minipage}
\caption{Histogram of the inclination angles of all computed models for $\beta$ Hydri in which $\chi^2$ is used as weight. See text for more details.}
\label{chiinclbh}
\end{figure}

For $\beta$~Hydri the degrees of the modes with highest power have been identified using the asymptotic relation \citep{tassoul1980} by \citet{bedding2007}. From the HARPS data that we have at our disposal we obtain five frequencies in common with \citet{bedding2007}, or one day aliases, for which we were able to analyse the line profile variations. For three common frequencies, \citet{bedding2007} determined $\ell$ = 0 and for the other two modes $\ell$ = 1 was found. We use these five frequencies for the line profile analysis. The results of the best fits to the observed line-profile variations obtained with the FPF method are shown in Figs.~\ref{ZAPbhydri1} and \ref{ZAPbhydri2}. We list the equivalent width, projected rotational velocity ($\upsilon \sin i$) and macro turbulence $\xi_{macro}$ for all four stars in Table~\ref{inppar}. The $\upsilon \sin i$ we obtain here for the line profile of $\beta$ Hydri is slightly higher than the independent determination by \citet{reiners2003}.

Inspection of the amplitude and phase distributions reveals clearly that for the radial modes the fits to the observed line-profile variations are consistent within their errors (see Fig.~\ref{ZAPbhydri1}). For the two modes with $\ell$ = 1 we fitted $-\ell \leq m \leq \ell$ and plot the fits for the zonal mode and modes with $m$ = 1 and $m$ = $-$1 for oscillations with frequencies 1033.1 $\mu$Hz and 1192.3 $\mu$Hz respectively (see left and centre panels of Fig.~\ref{ZAPbhydri2}). For $\nu$ = 1033.1 $\mu$Hz the fit with $m$ = 0 seems to be the best fit, while for $\nu$ = 1192.3 $\mu$Hz the fit with $m$ = $-$1 seems favourable over the $m$ = 0 fit. Nevertheless, for both frequencies the zonal and sectoral mode are consistent with the observations within the errors.

Because of the non-radial nature of the latter two modes, the line profile analysis depends on the inclination angle of the star. This inclination angle together with the projected rotational velocity leads to a surface rotational frequency ($\Omega$). For the inclination angle we take the interval defined by the weighted mean and standard deviation. The weight is defined as $w_k=\chi^2_0/\chi^2_k$, where $\chi^2_0$ is the $\chi^2$ of the best solution, see for more details \citet{desmet2009}. For this star, we find an inclination of 55 $\pm$ 17$^{\circ}$. An histogram of the weighted inclination is shown in Fig.~\ref{chiinclbh}. So for $\beta$ Hydri we find a surface rotational frequency ranging between 3.6 $\mu$Hz and 5.5 $\mu$Hz for inclination angles between 72$^{\circ}$ and 38$^{\circ}$ and a projected rotational velocity of 4.3 km\,s$^{-1}$ as obtained with FAMIAS.


\subsection{$\delta$~Eridani}

\begin{figure*}
\begin{minipage}{5.6cm}
\centering
\includegraphics[width=5.5cm]{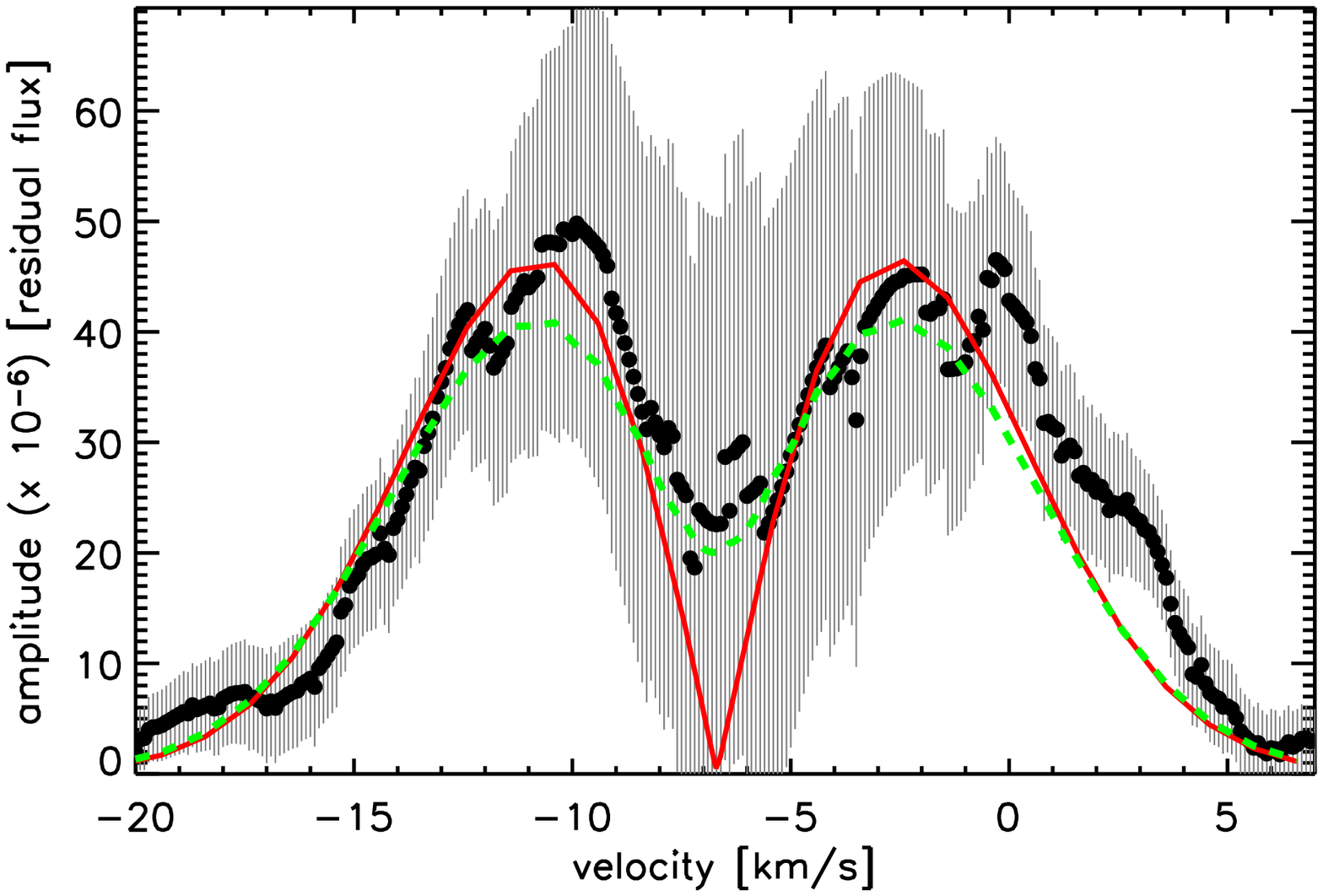}
\end{minipage}
\hfill
\begin{minipage}{5.6cm}
\centering
\includegraphics[width=5.5cm]{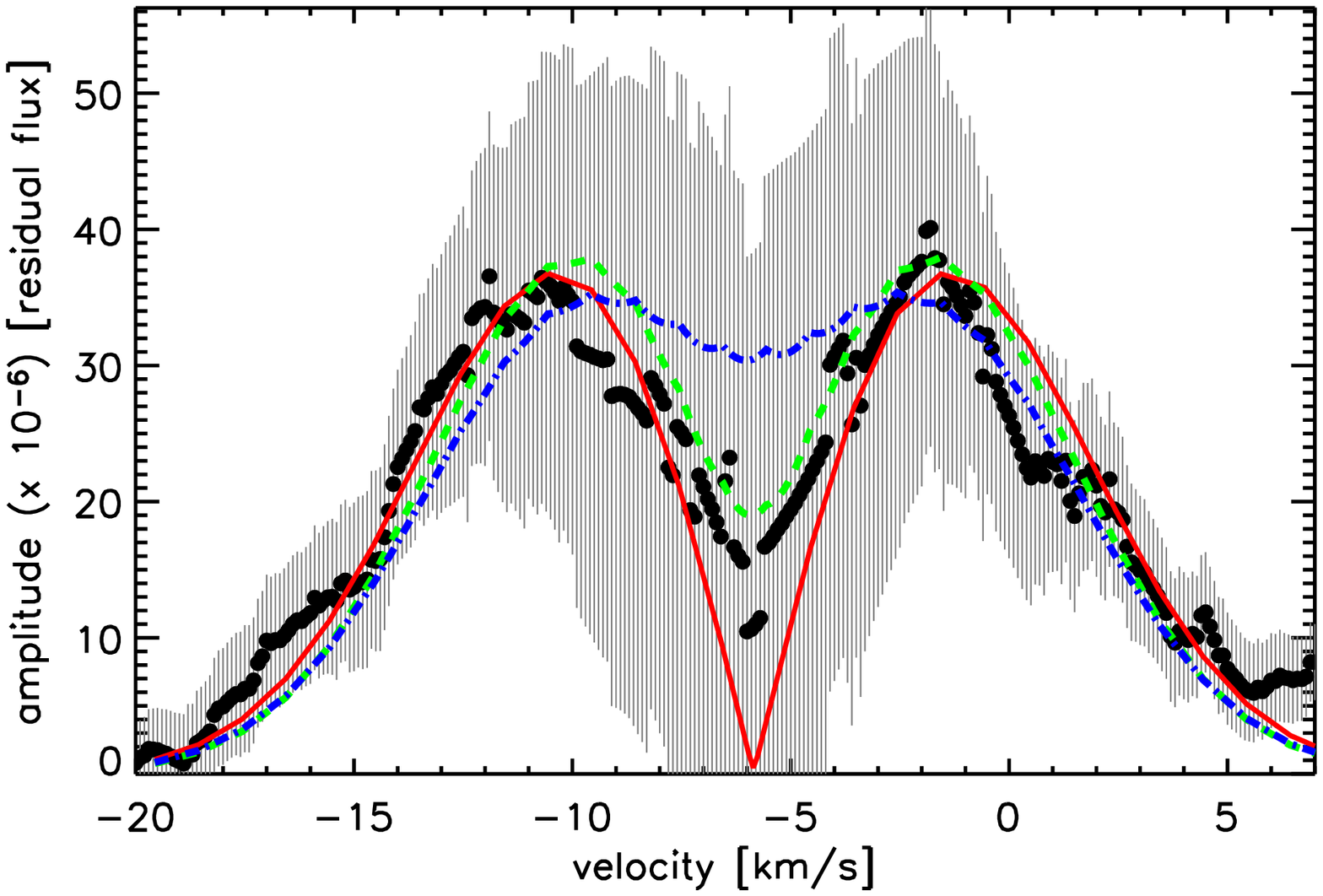}
\end{minipage}
\hfill
\begin{minipage}{5.6cm}
\centering
\includegraphics[width=5.5cm]{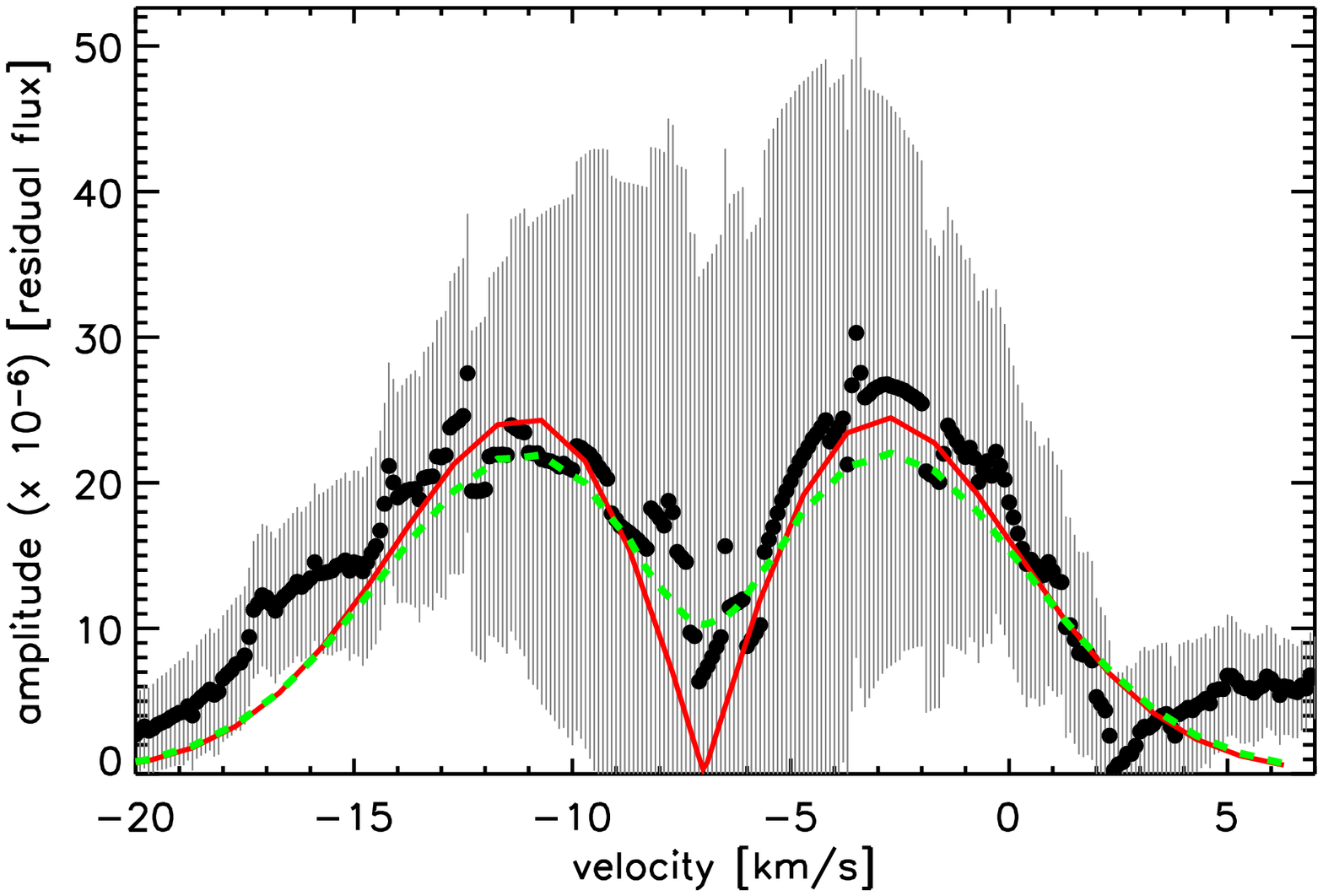}
\end{minipage}
\hfill
\begin{minipage}{5.6cm}
\centering
\includegraphics[width=5.5cm]{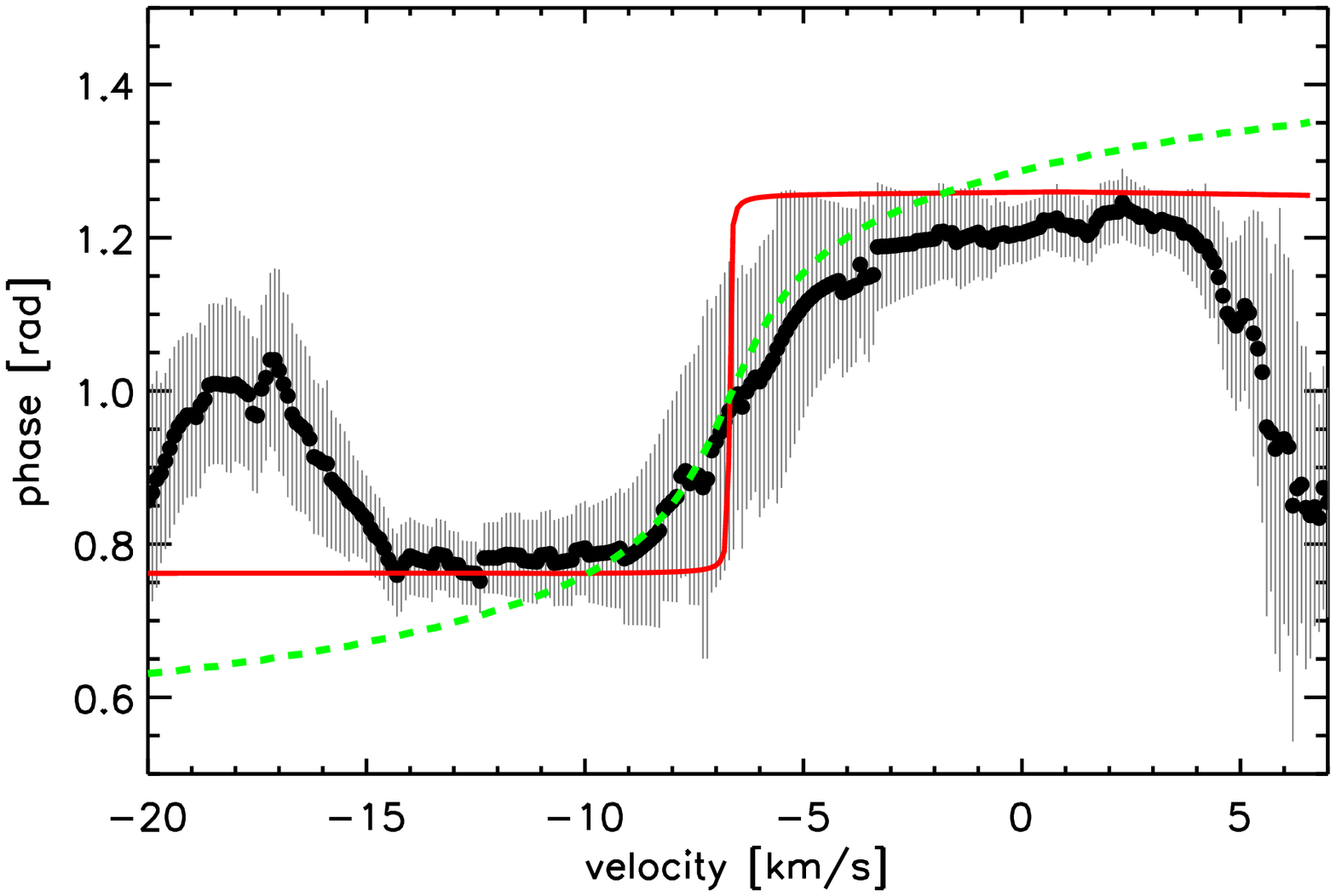}
\end{minipage}
\hfill
\begin{minipage}{5.6cm}
\centering
\includegraphics[width=5.5cm]{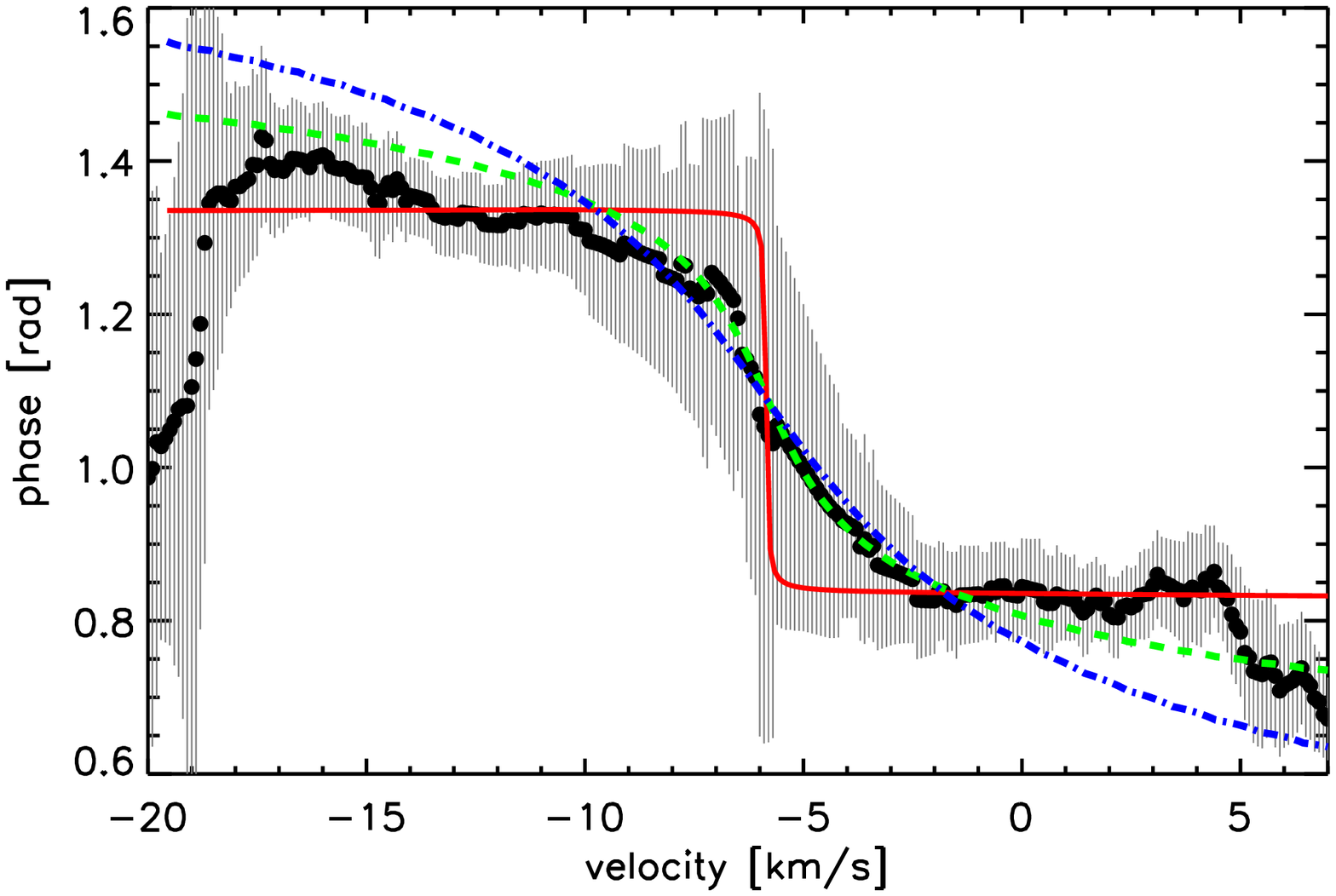}
\end{minipage}
\hfill
\begin{minipage}{5.6cm}
\centering
\includegraphics[width=5.5cm]{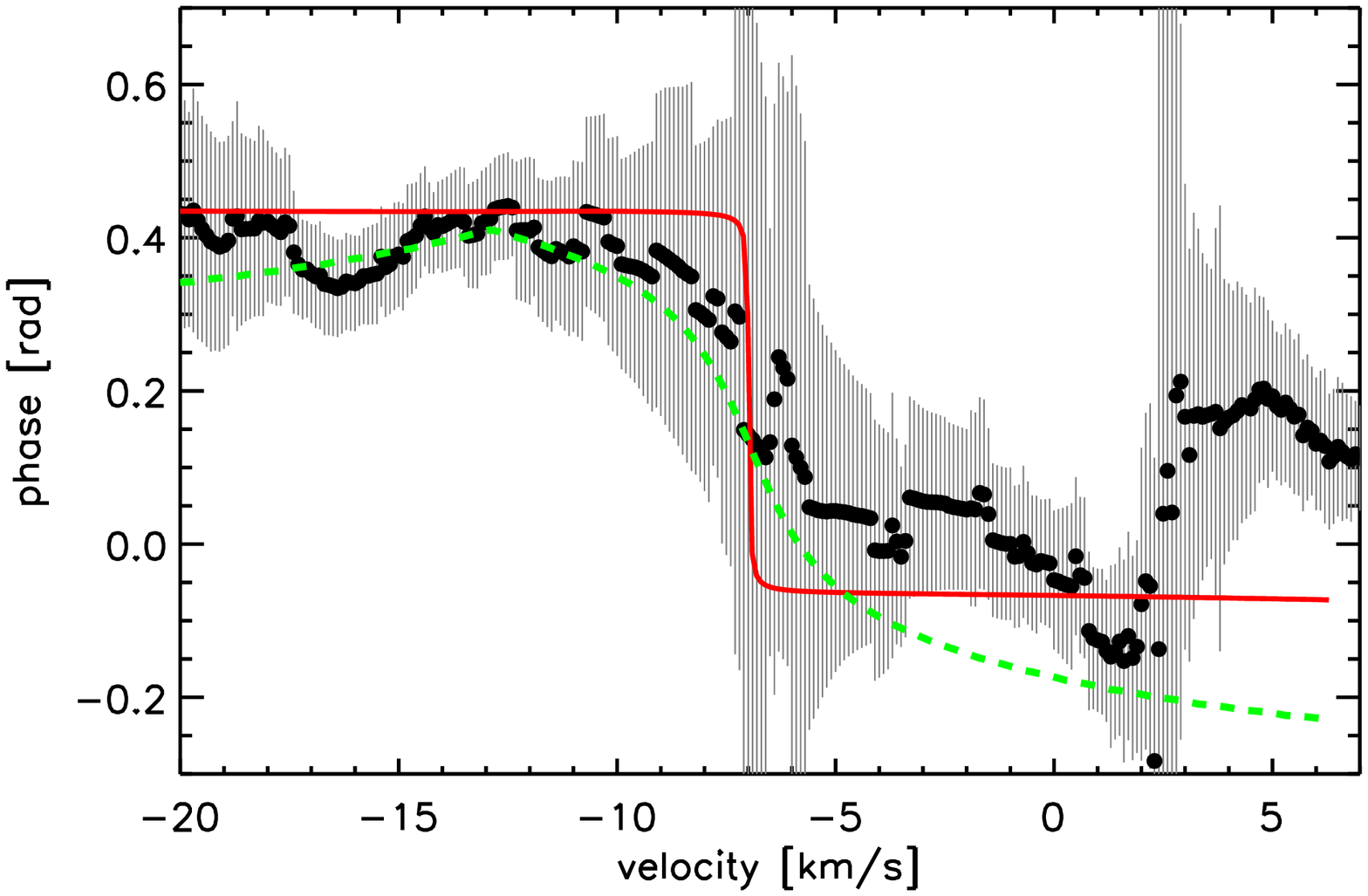}
\end{minipage}
\caption{Same as Fig.~\ref{ZAPbhydri1} for the significant frequencies of $\delta$~Eridani obtained from $\langle \mathrm{v} \rangle$: $\nu_{\langle \mathrm{v} \rangle}=58.4$~c\,d$^{-1}$ (675.8~$\mu$Hz) (left), $\nu_{\langle \mathrm{v} \rangle}=61.1$~c\,d$^{-1}$ (706.9~$\mu$Hz) (centre) and $\nu_{\langle \mathrm{v} \rangle}=57.5$~c\,d$^{-1}$ (665.9~$\mu$Hz) (right). The mean radial velocity of the star is approximately $-6.3$~km\,s$^{-1}$. The fits with $m$ = 0, $\pm$1, $\pm$ 2 are indicated with red solid, green dashed and blue dotted lines, respectively.}
\label{ZAPderi1}
\end{figure*}

\begin{figure}
\begin{minipage}{\linewidth}
\centering
\includegraphics[width=\linewidth]{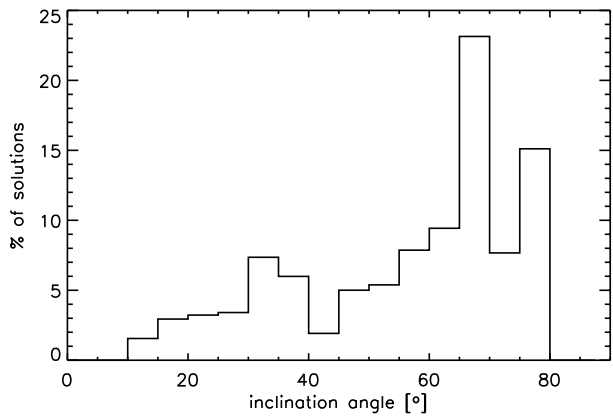}
\end{minipage}
\caption{Same as Fig.~\ref{chiinclbh} for $\delta$ Eridani.}
\label{chiinclde}
\end{figure}


\citet{carrier2002,carrier2003} analysed the radial velocity variations of $\delta$ Eridani from CORALIE observations and found evidence for 15 oscillation modes whose harmonic degrees could be identified using an \'echelle diagram. We obtain three frequencies in common (or 1 day aliases) in the variations of the first moment of the same data set, with enough signal-to-noise ratio to analyse the line-profile variations. The harmonic degrees of all three oscillation frequencies are known and used as input for the present line-profile analysis. The results of this analysis are listed in Table~\ref{inppar} and shown in Fig.~\ref{ZAPderi1}.

The most dominant mode we obtain from the first moment has a frequency of 675.8 $\mu$Hz and was identified by \citet{carrier2002,carrier2003} as a radial mode. Our fit of a radial mode, is indeed consistent with the data within errors. The fact that the 'centre' part of the amplitude distribution does not drop to 0 is an indication of the presence of a non-radial mode. Indeed a fit with $(\ell,m)$ = (1,$-$1) does provide a better fit to the amplitude distribution and the centre part of the phase distribution (see left panels of Fig.~\ref{ZAPderi1}). 

The second frequency is identified by \citet{carrier2002,carrier2003} as an $\ell$ = 2 mode and from our present analysis we see that $(\ell,m)$ = (2,1) seems the most likely mode identification (see centre panels of Fig.~\ref{ZAPderi1}). The third frequency is consistent with $\ell$ = 1 as determined by \citet{carrier2002,carrier2003}, but we are not able to discriminate between $m$ = 0 or $m$ = $-$1 due to the lower signal to noise ratio, which mostly influences the phase distribution of this frequency (see right panels of Fig.~\ref{ZAPderi1}). 

Also for this star we investigated the weighted mean and standard deviation of all inclination angles of the computed synthetic profiles (see Fig.~\ref{chiinclde} for an histogram in which the weights are taken into account) to compute the surface rotational frequency. The inclination angle is 57 $\pm$ 19$^{\circ}$, which results in a range of surface rotational frequencies between 3.1 and 4.9 $\mu$Hz at inclination angles ranging from 76 to 38$^{\circ}$ and a projected rotational velocity of 4.9 km\,s$^{-1}$.

\subsection{$\epsilon$~Ophiuchi}


\begin{figure*}
\begin{minipage}{5.6cm}
\centering
\includegraphics[width=5.5cm]{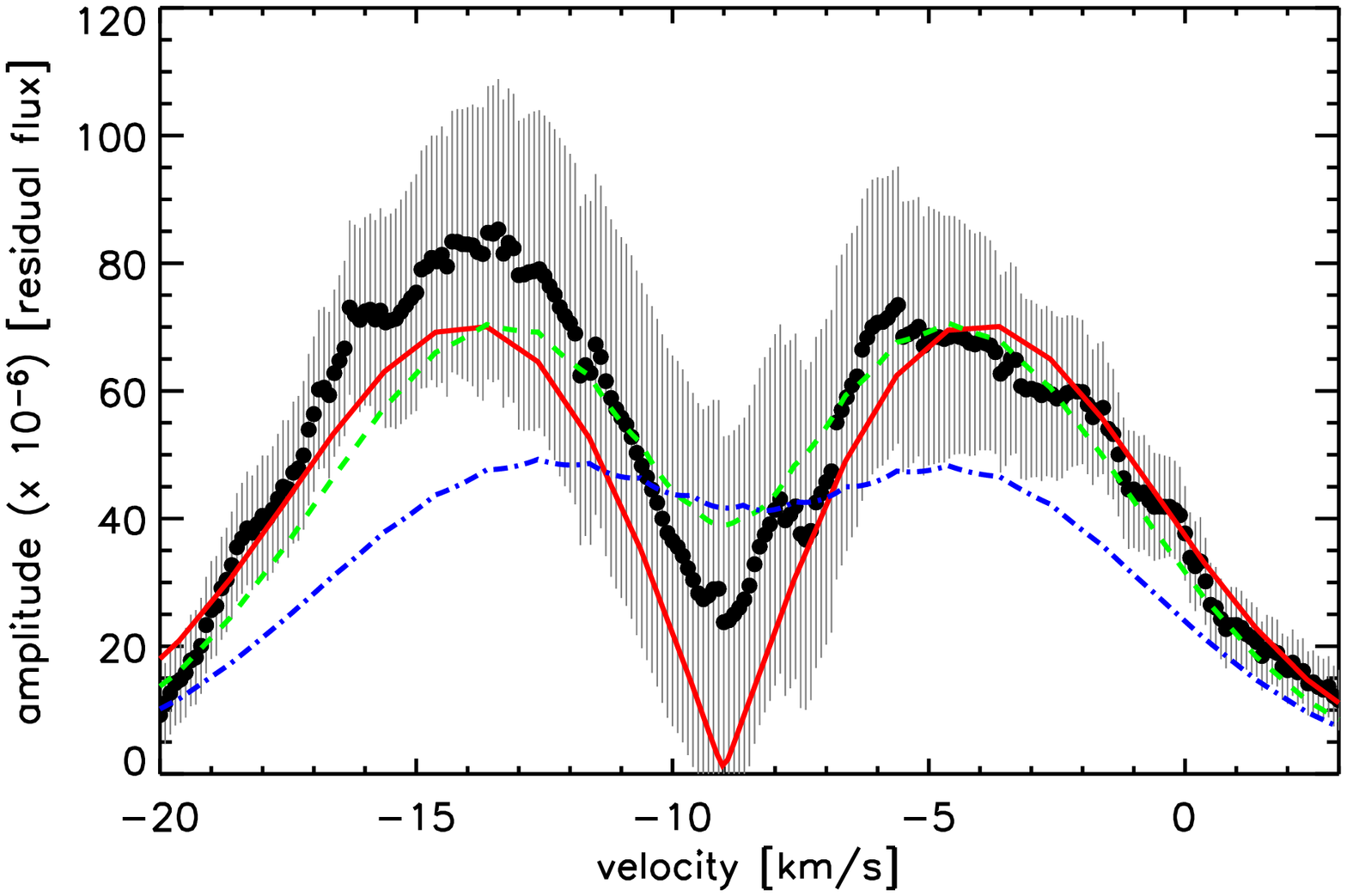}
\end{minipage}
\hfill
\begin{minipage}{5.6cm}
\centering
\includegraphics[width=5.5cm]{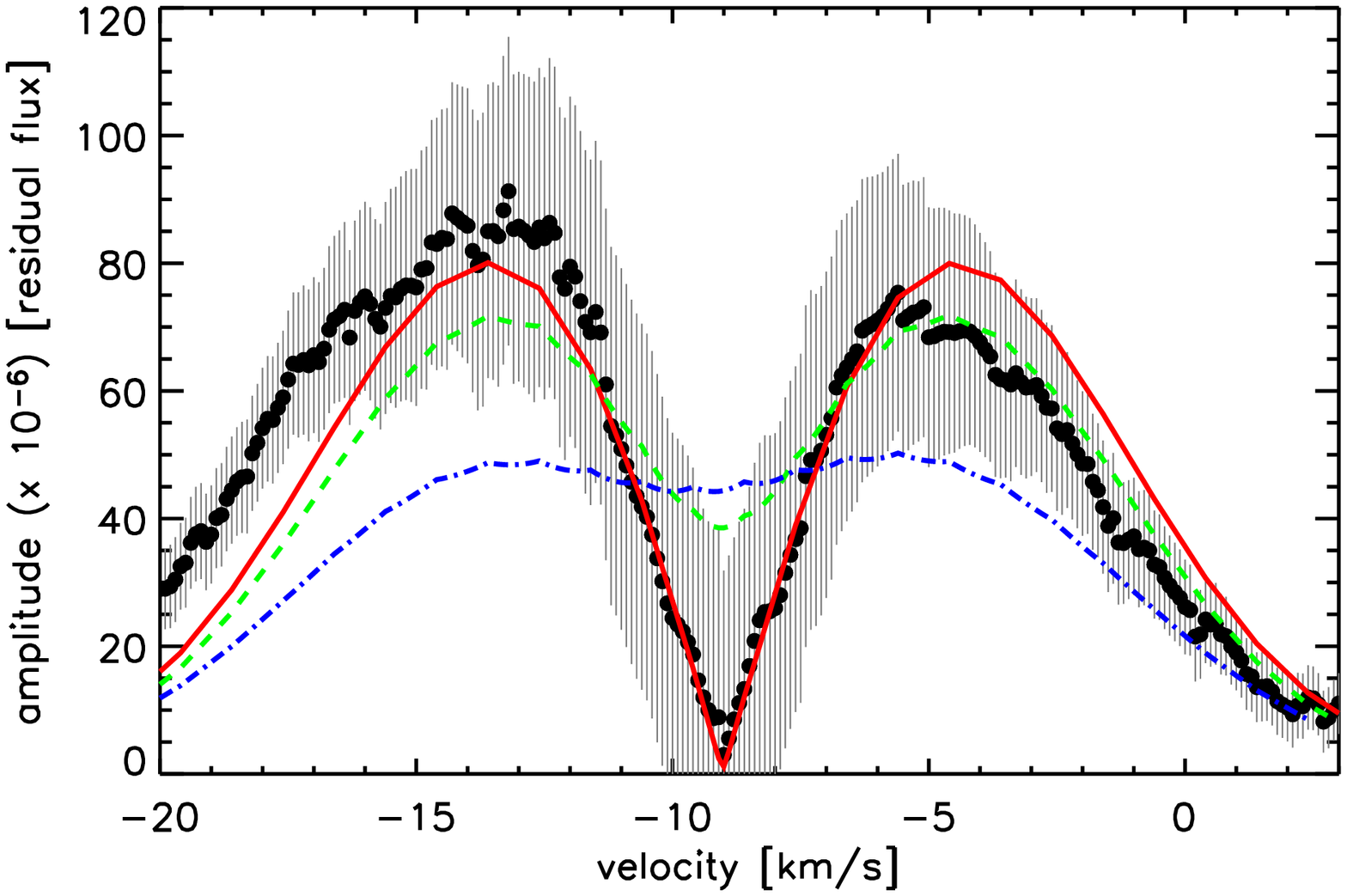}
\end{minipage}
\hfill
\begin{minipage}{5.6cm}
\centering
\includegraphics[width=5.5cm]{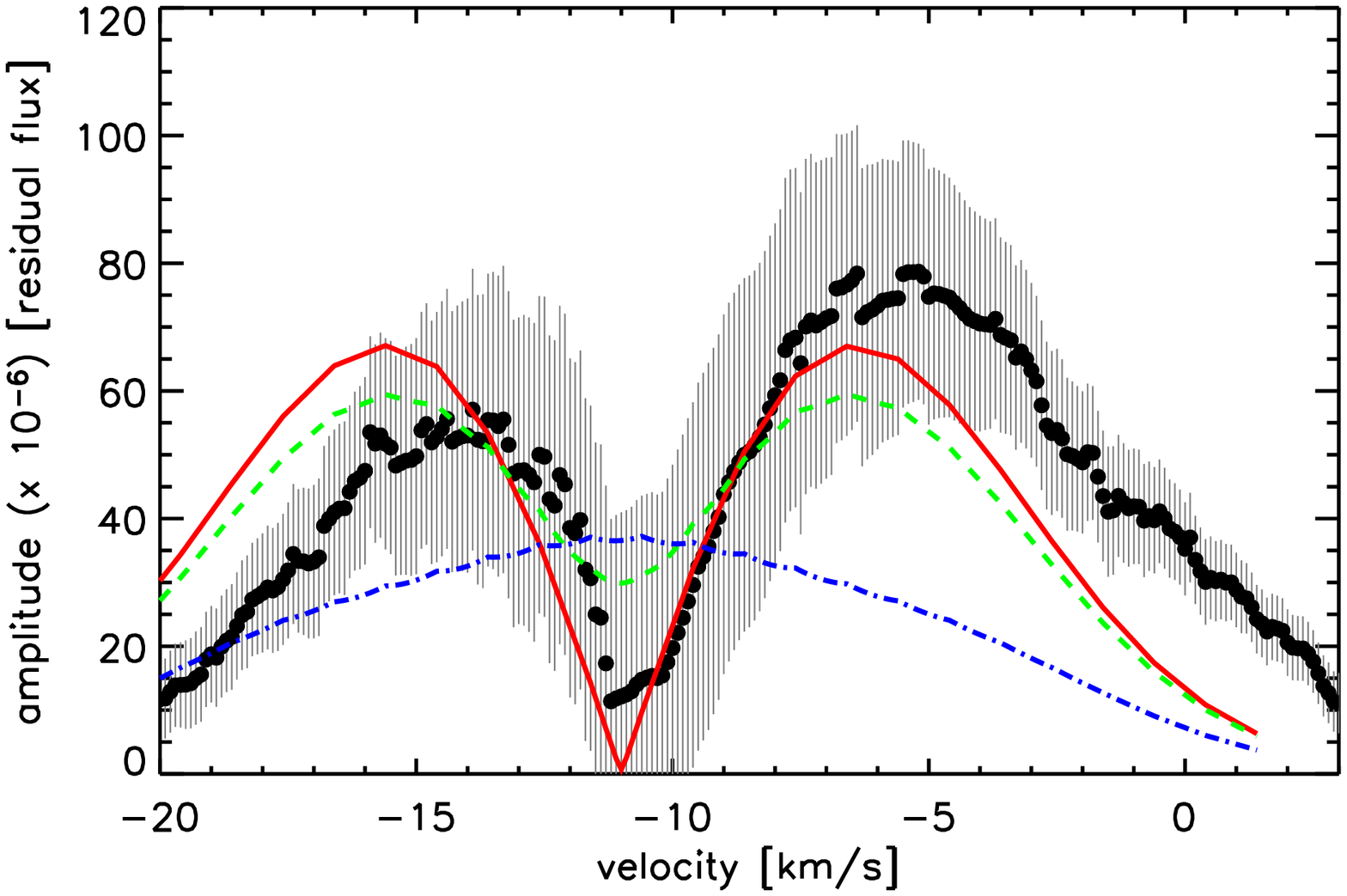}
\end{minipage}
\hfill
\begin{minipage}{5.6cm}
\centering
\includegraphics[width=5.5cm]{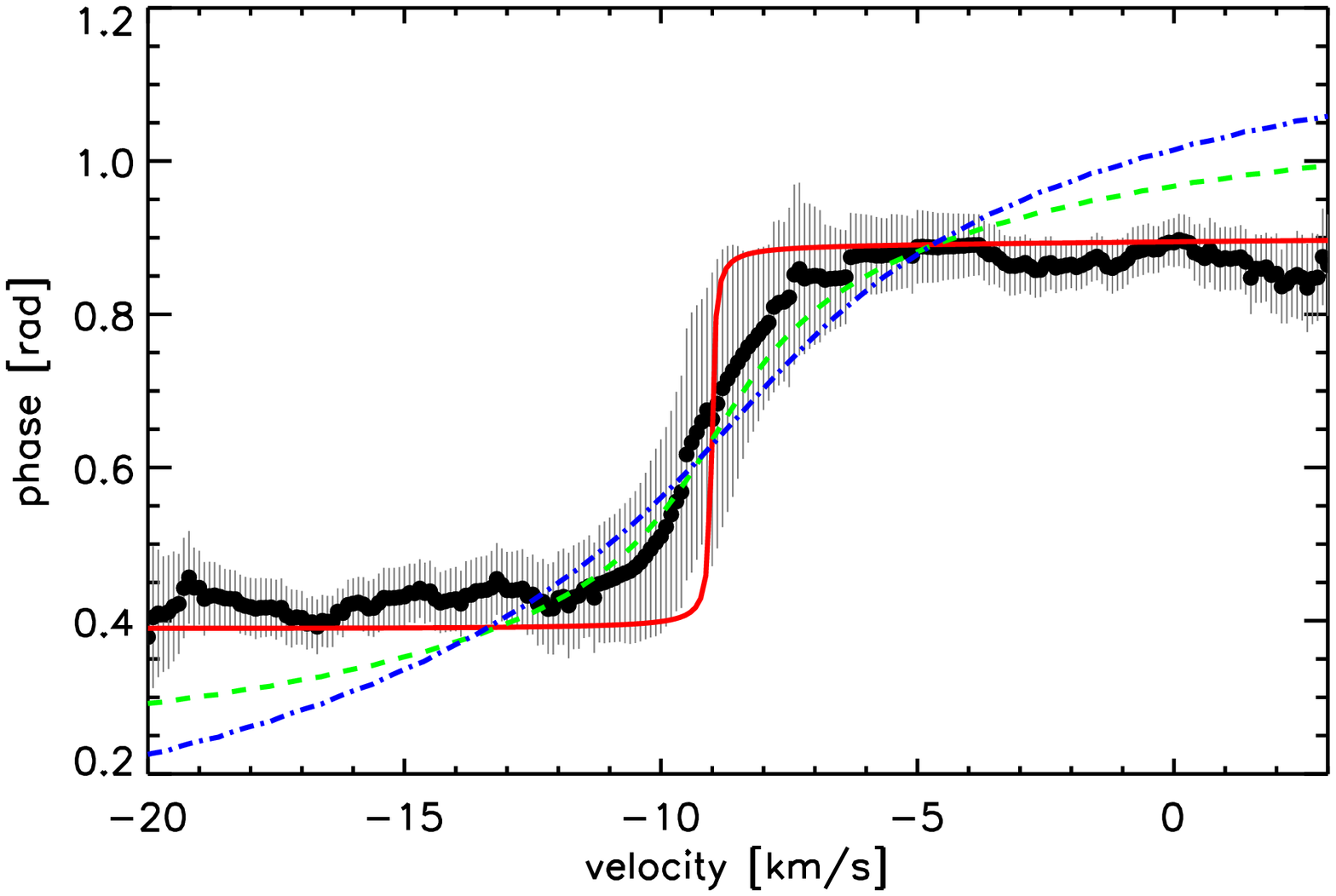}
\end{minipage}
\hfill
\begin{minipage}{5.6cm}
\centering
\includegraphics[width=5.5cm]{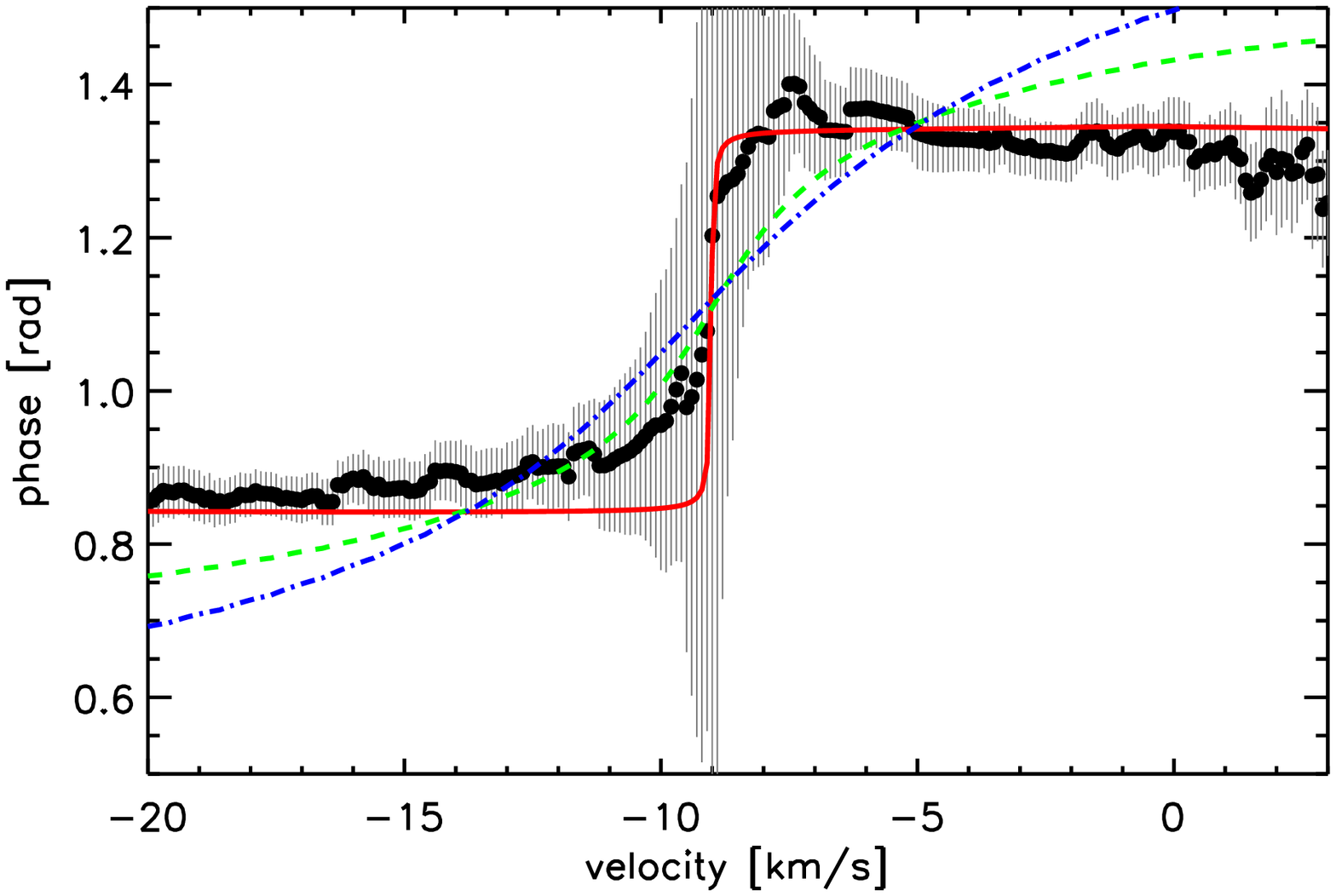}
\end{minipage}
\hfill
\begin{minipage}{5.6cm}
\centering
\includegraphics[width=5.5cm]{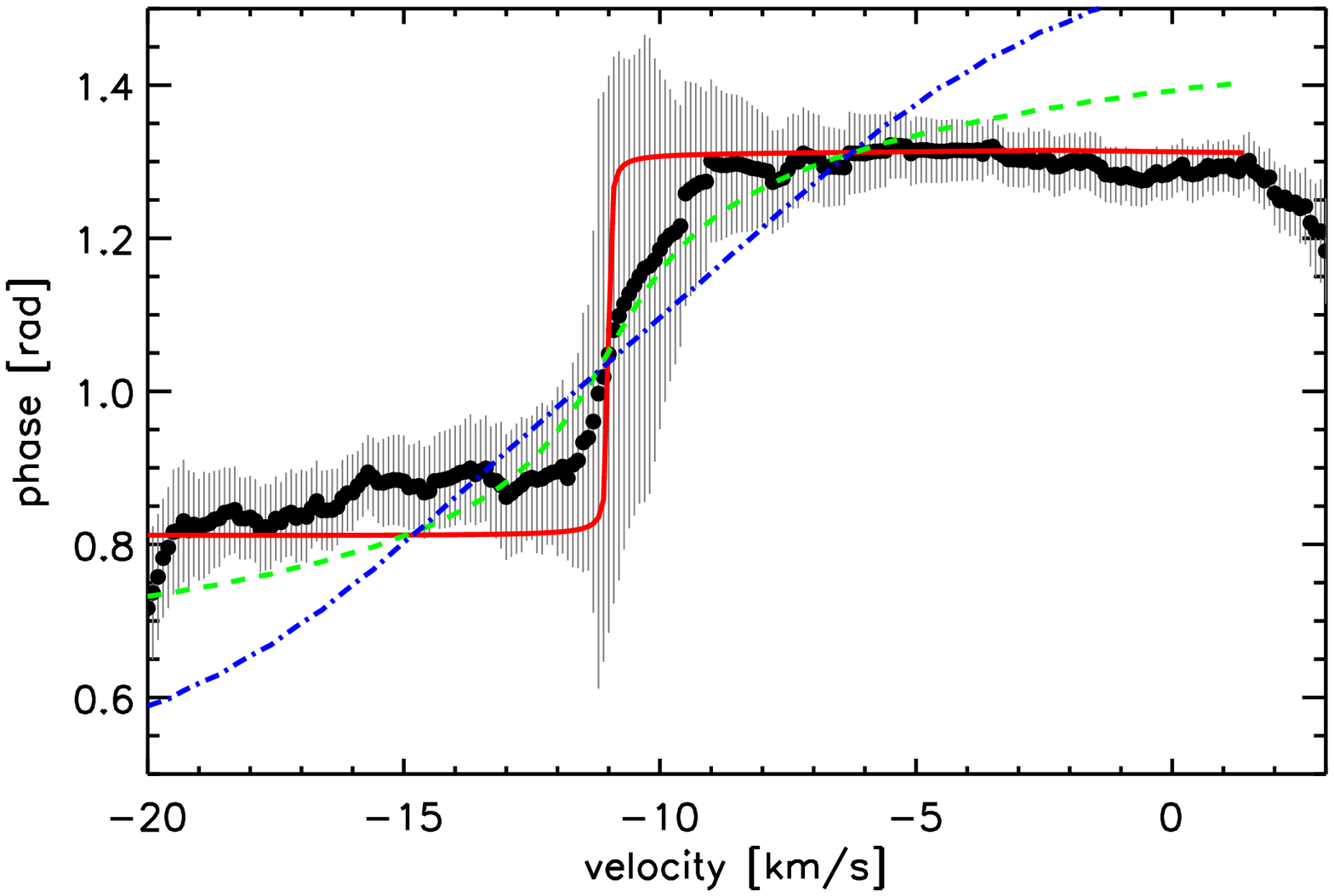}
\end{minipage}
\caption{Same as Fig.~\ref{ZAPderi1} for the significant frequencies of $\epsilon$~Ophiuchi obtained from $\langle \mathrm{v} \rangle$: $\nu_{\langle \mathrm{v} \rangle}=5.03$~c\,d$^{-1}$ (58.2~$\mu$Hz) (left), $\nu_{\langle \mathrm{v} \rangle}=4.65$~c\,d$^{-1}$ (53.8~$\mu$Hz) (centre) and $\nu_{\langle \mathrm{v} \rangle}=6.46$~c\,d$^{-1}$ (74.8~$\mu$Hz) (right). The mean radial velocity of the star is approximately $-9.4$~km\,s$^{-1}$.}
\label{ZAPepsoph}
\end{figure*}

\begin{figure}
\begin{minipage}{\linewidth}
\centering
\includegraphics[width=\linewidth]{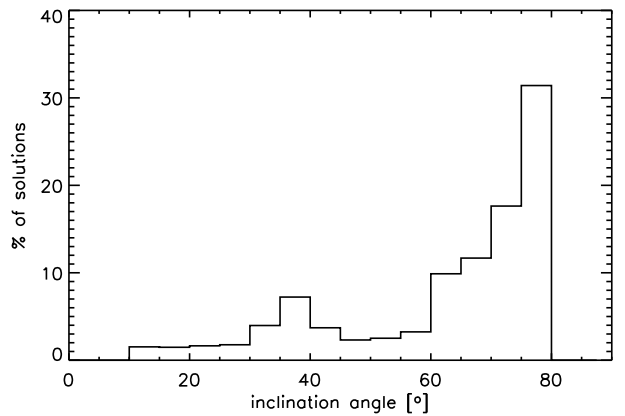}
\end{minipage}
\caption{Same as Fig.~\ref{chiinclbh} for $\epsilon$ Ophiuchi.}
\label{chiincleo}
\end{figure}

\citet{deridder2006} and \citet{barban2007} interpreted the frequencies observed in $\epsilon$~Ophiuchi, obtained from radial velocities derived from CORALIE and ELODIE and from photometric space data (MOST) respectively, as radial modes. Based on the combined data set \citet{kallinger2008} interpreted the frequencies of $\epsilon$~Ophiuchi as both radial and non-radial oscillation modes. They performed mode identification by comparing the observed frequencies with those of theoretical models to derive the degree of the modes. Hence this mode identification is model dependent. From the (same) CORALIE data we have at our disposal we obtain three frequencies of which only for one frequency the harmonic degree of the oscillation has already been identified by \citet{kallinger2008}. The results of the best fits we obtain with the FPF method are listed in Table~\ref{inppar} and shown in Fig.~\ref{ZAPepsoph}.

A first inspection of the amplitude and phase distribution immediately shows that $m$ = $-$2 provides for all frequencies the least likely mode identification.

For the 1$^{st}$ frequency we find that both fits for $m$ = 0 as well as $m$ = $-$1 are within the errors consistent with the amplitude distribution for the full CCF, where the $m$ = $-$1 fit is superior in the centre of the line (see left top panel of Fig.~\ref{ZAPepsoph}).  This is also the case in the centre of the phase distribution, where the $m$ = $-$1 fit matches the observations closely, while the $m$ = 0 fit lays outside the error bars. In the wings of the CCF the opposite is the case, i.e., the $m$ = 0 fit matches the observations and the $m$ = $-$1 fit falls outside the error bars. Thus from the line centre we would conclude that $m$= $-$1 and from the line wings that $m$ = 0. Both are consistent with \citet{kallinger2008} and only the latter is consistent with \citet{deridder2006} and \citet{barban2007}. 

The observed amplitude distribution of the 2$^{nd}$ frequency shows a clear dip in the centre which is matched very closely by the $m$ = 0 fit, while the $m$ = $-$1 fit lays just within the error in this region (centre top panel of Fig.~\ref{ZAPepsoph}). The $m$ = 0 fit also matches the phase distribution best in both the centre and the wings, which implies this is the most likely mode identification.

The asymmetry in the observed amplitude distribution of the 3$^{rd}$ frequency hampers a good fit, but in the line centre both $m$ = 0 and $m$ = $-$1 fall well within the error bars (right top panel of Fig.~\ref{ZAPepsoph}). The centre part of the phase distribution is best fitted with the $m$ = $-$1 fit while the line wings are again best fitted with an $m$ = 0 fit. Thus here, the centre of the line favours a $m$ = $-$1 mode identification, but $m$ = 0 is also possible.

For this star the preferred inclination angles, computed from the weighted mean and standard deviation, all lay in the range 63 $\pm$ 18$^{\circ}$, which, together with the projected rotational velocity of 5.7 km\,s$^{-1}$, results in an surface rotational frequency ranging between 0.8 and 1.1 $\mu$Hz.

\subsection{$\eta$~Serpentis}

\begin{figure*}
\begin{minipage}{5.6cm}
\centering
\includegraphics[width=5.5cm]{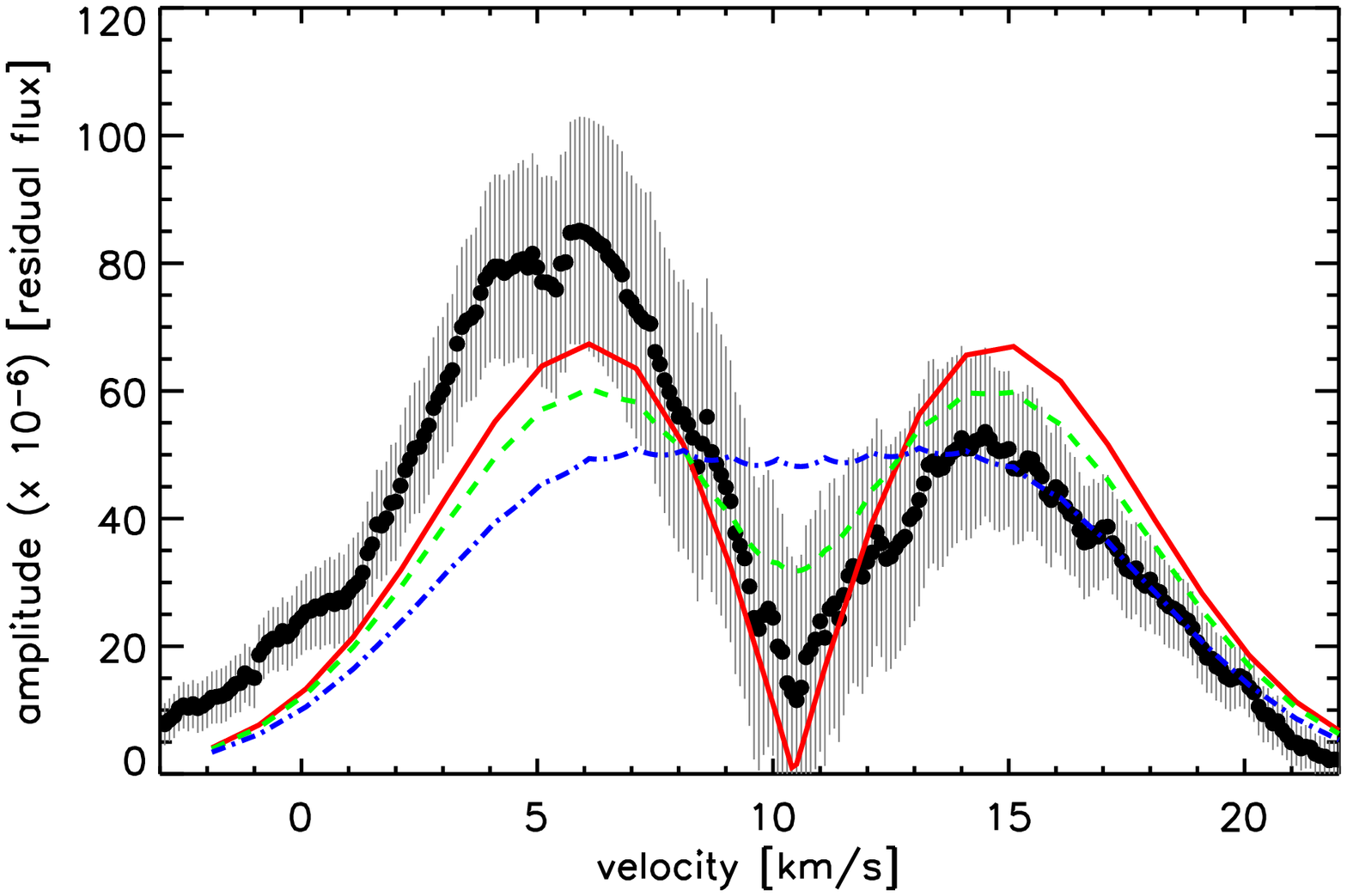}
\end{minipage}
\hfill
\begin{minipage}{5.6cm}
\centering
\includegraphics[width=5.5cm]{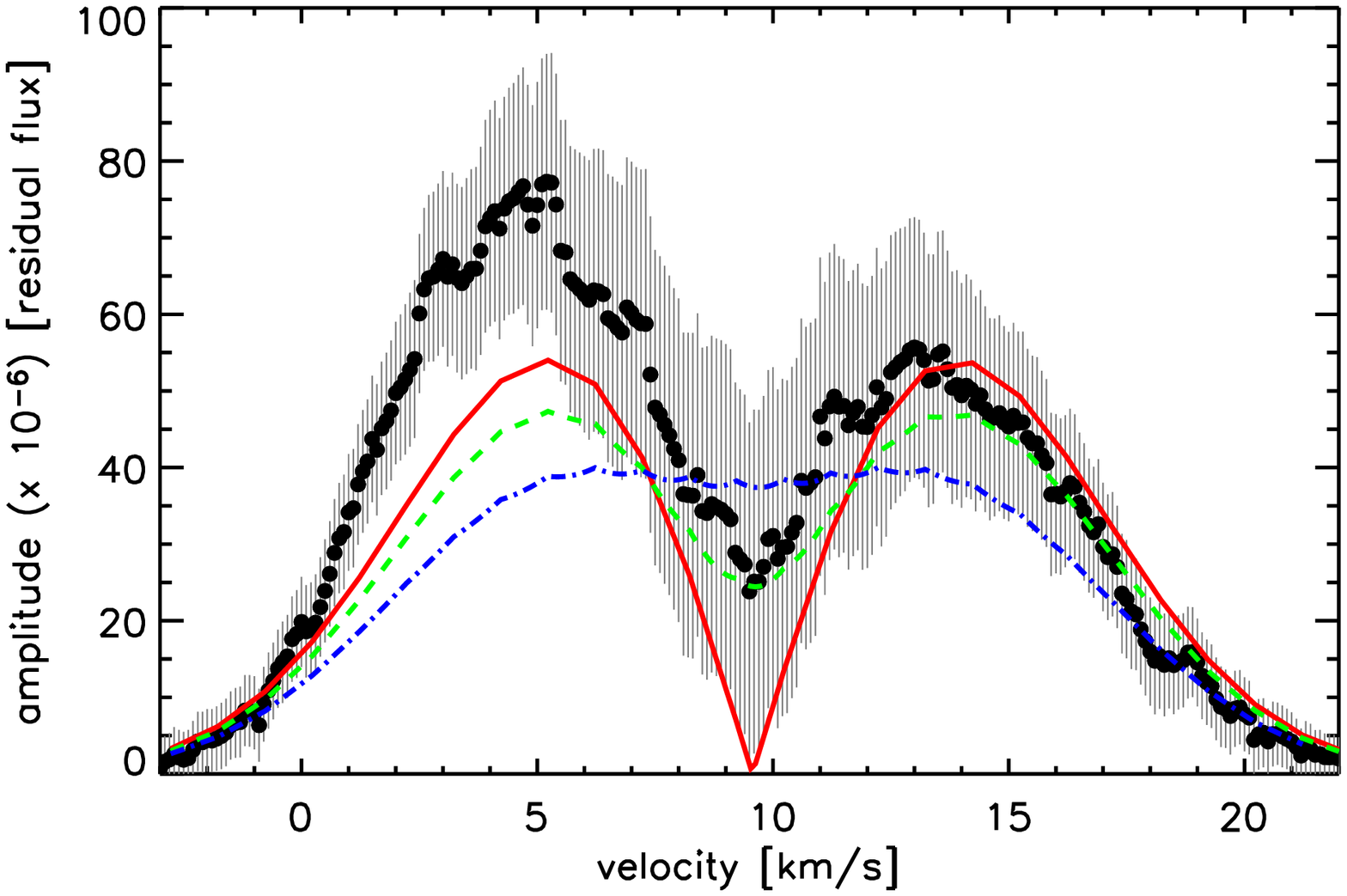}
\end{minipage}
\hfill
\begin{minipage}{5.6cm}
\centering
\includegraphics[width=5.5cm]{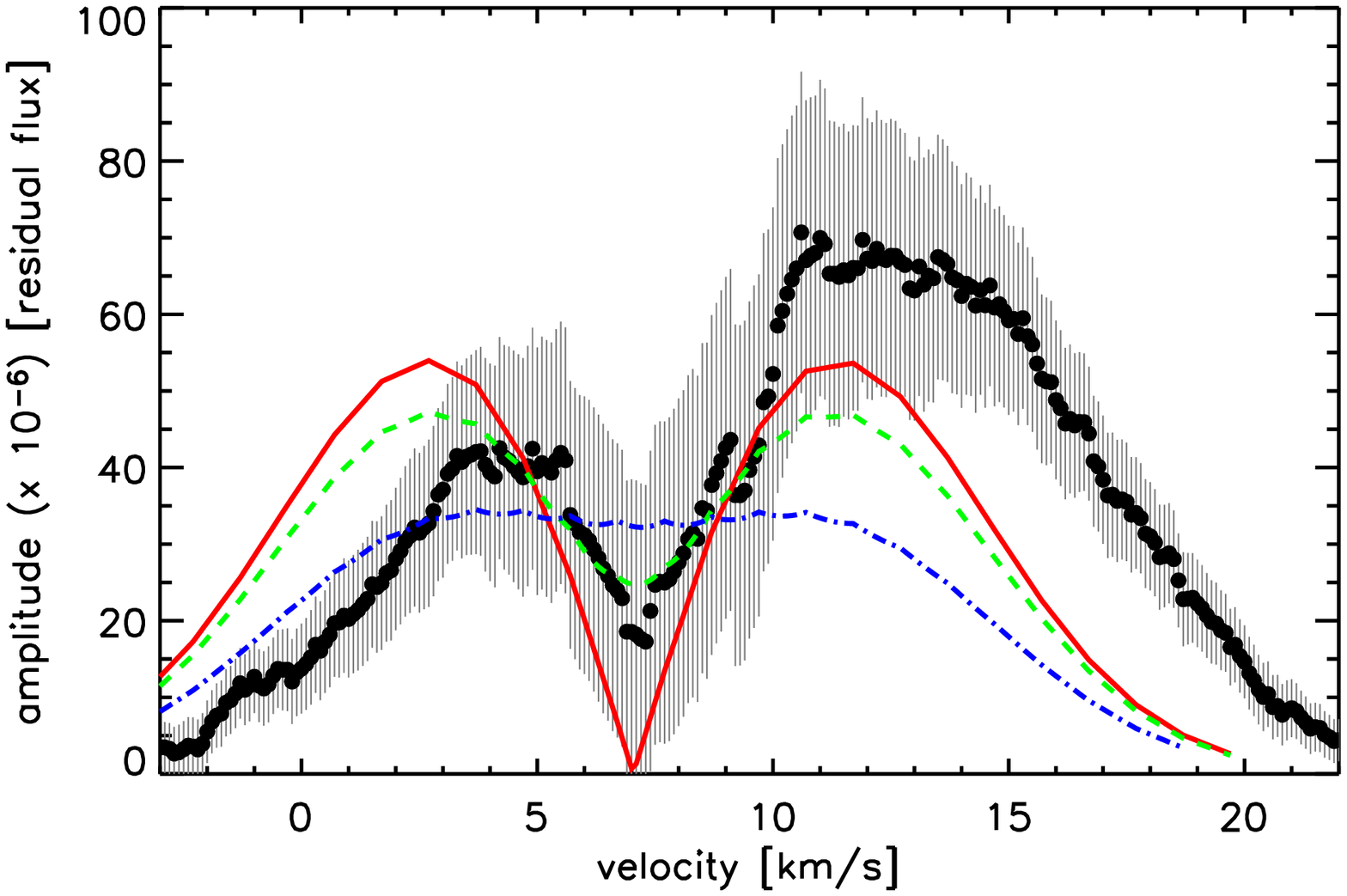}
\end{minipage}
\hfill
\begin{minipage}{5.6cm}
\centering
\includegraphics[width=5.5cm]{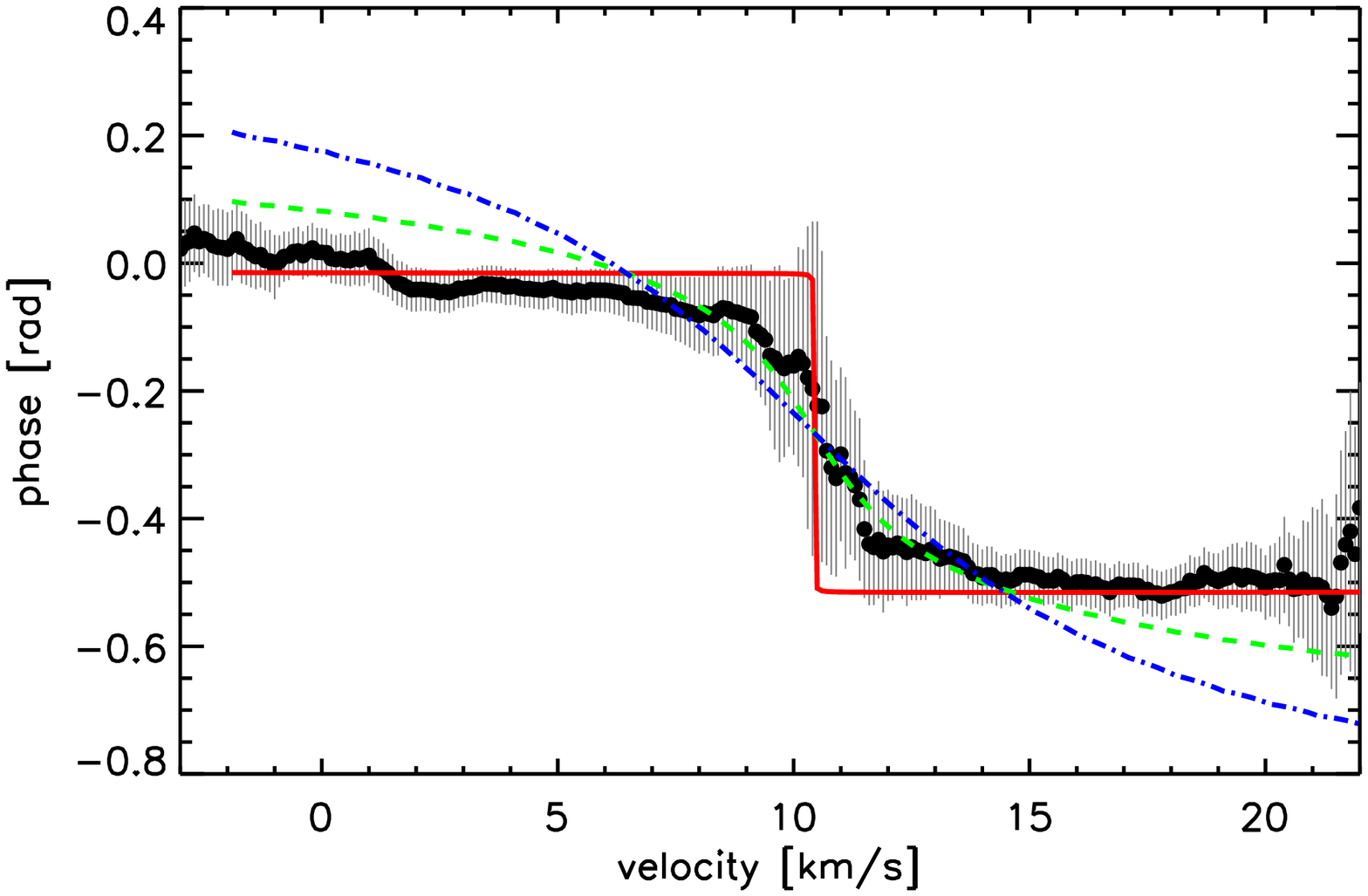}
\end{minipage}
\hfill
\begin{minipage}{5.6cm}
\centering
\includegraphics[width=5.5cm]{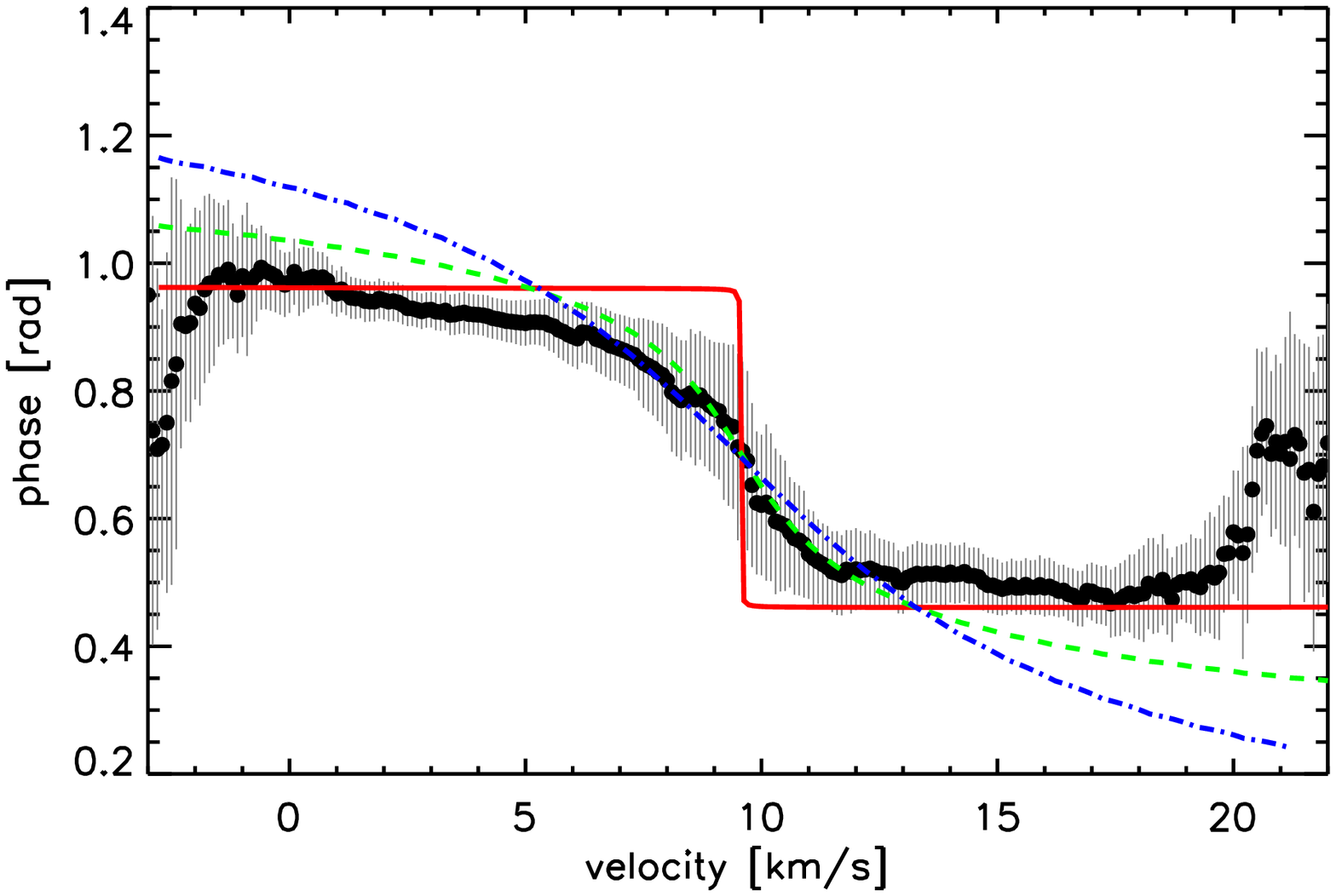}
\end{minipage}
\hfill
\begin{minipage}{5.6cm}
\centering
\includegraphics[width=5.5cm]{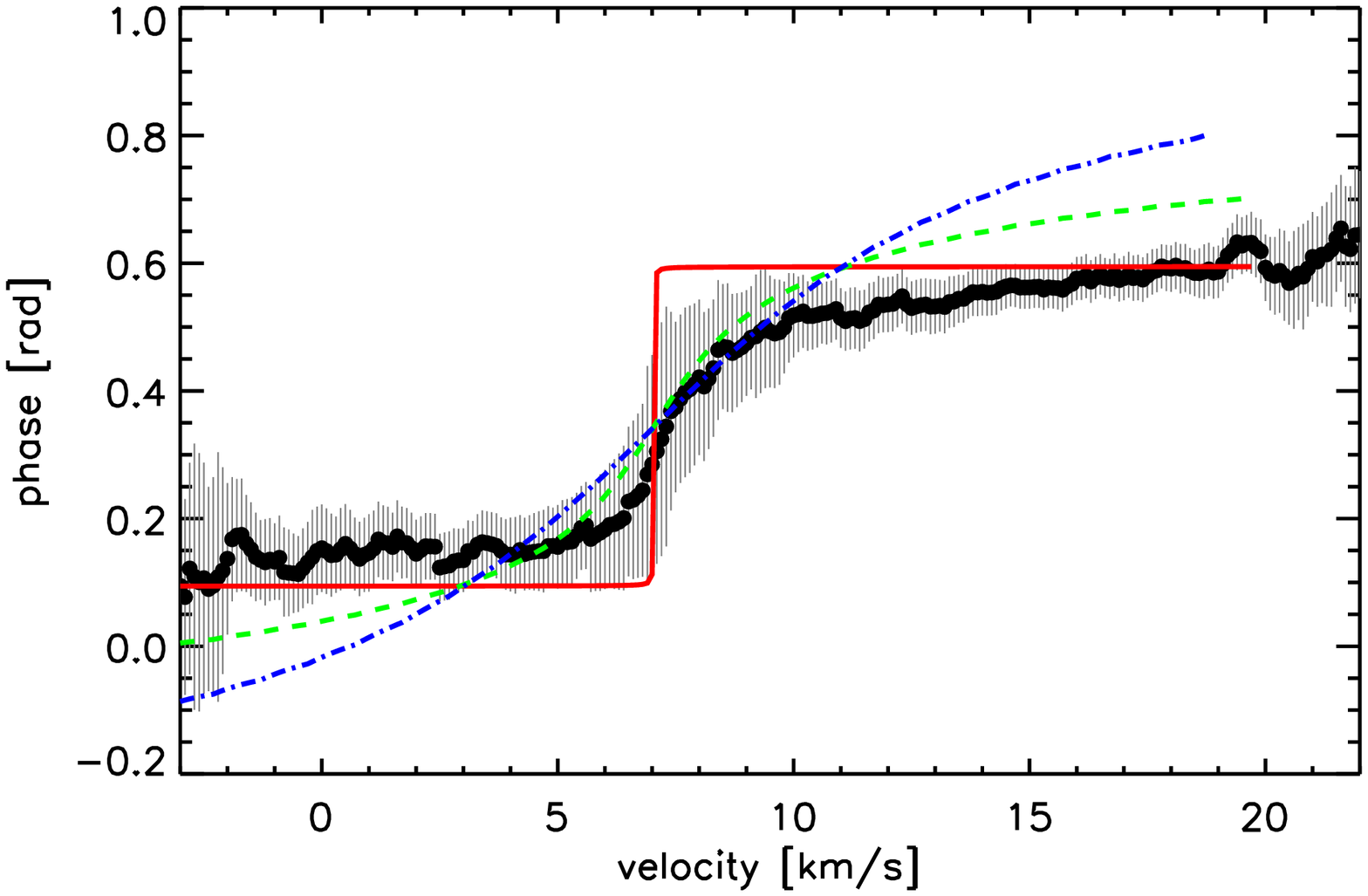}
\end{minipage}
\caption{Same as Fig.~\ref{ZAPderi1} for the significant frequencies of $\eta$~Serpentis obtained from $\langle \mathrm{v} \rangle$: $\nu_{\langle \mathrm{v} \rangle}=11.74$~c\,d$^{-1}$ (135.9~$\mu$Hz) (left), $\nu_{\langle \mathrm{v} \rangle}=9.48$~c\,d$^{-1}$ (109.7~$\mu$Hz) (centre) and $\nu_{\langle \mathrm{v} \rangle}=10.38$~c\,d$^{-1}$ (120.1~$\mu$Hz) (right). The mean radial velocity of the star is approximately $9.4$~km\,s$^{-1}$.}
\label{ZAPetaser}
\end{figure*}

After confirmation that we can compare the line-profile variations from observations and synthetic spectra quantitatively, we have analysed the CCFs of $\eta$~Serpentis. The results are shown in Table~\ref{inppar} and shown in Fig.~\ref{ZAPetaser}. For these stars we see again that $m$ = $-$2 and $m$ = 2 are the least favoured mode identification for all frequencies.

For $\eta$~Serpentis all amplitude distributions are asymmetric and therefore none of the fits are consistent in the line wings, while in the centre both $m$ = 0  and $m$ = $-$1 or $m$ = 1 are consistent with the observed distribution. In the centre of the phase distribution the $m$ = 1(1$^{st}$ and 2$^{nd}$ frequency) and $m$ = $-$1 (3$^{rd}$ frequency) fits match the observations closely, while the $m$ = 0 fits are just within the errors (1$^{st}$ and 2$^{nd}$ frequency) or even outside the error bars (3$^{rd}$ frequency). The phase distributions in the CCF wings are best matched by the $m$ = 0 fits. For this star the line centres favour an $m$ = 1(1$^{st}$ and 2$^{nd}$ frequency) or $m$ = $-$1 (3$^{rd}$ frequency) mode identification, while $m$ = 0 cannot be ruled out because of the better fits in the CCF wings.

\begin{figure}
\begin{minipage}{\linewidth}
\centering
\includegraphics[width=\linewidth]{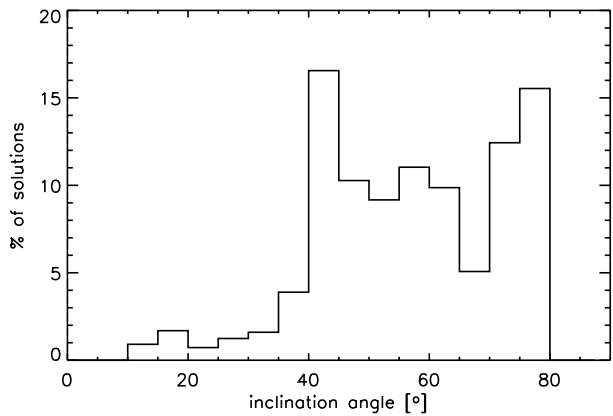}
\end{minipage}
\caption{Same as Fig.~\ref{chiinclbh} for $\eta$ Serpentis.}
\label{chiincles}
\end{figure}

The inclination angles of the synthetic line profiles fitted to the data have a weighted mean value of 57$^{\circ}$ with a standard deviation of 16$^{\circ}$ (see Fig.~\ref{chiincles} for the histogram). This range of inclination angles together with the projected rotational velocity determined with FAMIAS (Table~\ref{inppar}) results in a surface rotational frequency between 1.5 and 2.1 $\mu$Hz.

\section{Discussion and conclusions}
To take the analysis of line-profile variations performed by \citet{hekker2006} one step further, i.e., from a qualitative analysis to a quantitative analysis, it was necessary to apply a number of specific corrections to the cross-correlation functions. With these corrections we removed profiles with outliers at any velocity across the line profile. These outliers seem to have caused the large, single peaked amplitude profiles and low phase changes of the line-profile variations, most clearly seen in the dominant frequencies of $\epsilon$ Ophiuchi and $\eta$ Serpentis, see Figs.~6 and 7 of \citet{hekker2006}. \citet{hekker2006} could match these amplitude profiles qualitatively with $m$ = 2 modes, but the low phase change over the profile could not be reproduced from synthetic line profiles, which hampered a definite mode identification. The corrections applied here altered the amplitude and phase distributions and on these corrected distributions we could perform a quantitative analysis.

We have been able to analyse line-profile variations in the corrected CCFs of four evolved stars and identify the azimuthal order for the frequencies detected in the first moments with an uncertainty of $\pm$1 as is usually the case in spectroscopic mode identification with the adapted method. From the synthetic line profile fitting of non-radial modes as performed with FAMIAS, we could also determine a range of inclination angles for the stars. These together with the projected rotational velocity provided us with the surface rotational frequencies, which are all shown in Table~\ref{srv}.

The uncertainty in the mode identification is mostly due to different favourable identifications in the line wings and in the line centres. Data with higher signal-to-noise ratio, i.e., lower noise levels, than we have at hand could improve the mode identification significantly. Firstly, this would decrease the asymmetry in the observed amplitude distributions for modes with $m$ = 0, as seen from simulations. Secondly, lower noise values could reduce the errors on the amplitude and phase distributions which will allow us to better distinguish between fits with different $m$ values. Nevertheless it is possible that higher signal-to-noise ratio of the data will not improve the fits to the phase distribution. It is clear from the lower panels of Figs.~\ref{ZAPderi1}, \ref{ZAPepsoph} and \ref{ZAPetaser} that for $m$ $\neq$ 0 the phase difference in the wings changes gradually for the synthetic profiles, while this is less so for the observed profiles. We tested whether this discrepancy was due to the fact that we neglect temperature effects on the equivalent width, but taking those into account did not improve the fitting considerably.

For $\beta$~Hydri the azimuthal orders obtained from the line-profile analysis are compatible with the harmonic degree of the modes determined previously from radial velocity measurements and asymptotic frequency relations by \citet{bedding2007}. Three modes with $\ell$ = 0 and one mode with $\ell$ = 1 are compatible with $m$ = 0, while for the fifth frequency we know $\ell$ =1 and $m$ = $-$1 is clearly favourable in this case.

 The mode identifications for $\delta$ Eridani are also consistent with the results from \citet{carrier2002,carrier2003}. Although the present method seems to point to a non-radial mode for the dominant frequency instead of a radial as claimed by \citet{carrier2002,carrier2003}. Our identification would change the degrees assigned to the different ridges in the \'echelle diagram, but we do not have enough frequencies for which we could apply the current method to make any firm statements.

For $\epsilon$~Ophiuchi there is one mode in common between the present work and the frequency model fitting performed by \citet{kallinger2008}. The identifications for this non-radial oscillation mode are also in agreement. As we find two possible values of the azimuthal order for all frequencies, our analysis  could also be in agreement with the radial mode interpretation favoured by \citet{deridder2006} and \citet{barban2007}.

The confirmation that we can compare line-profile variations of observations and synthetic profiles quantitatively led us to also analyse $\eta$~Serpentis. Also for this stars we were able to obtain the azimuthal orders for three modes with an uncertainty of $\pm$ 1 and an indication of the surface rotational frequency. These identifications and the surface rotational frequency are less accurate due to lower signal to noise in the amplitude and phase diagrams which are therefore more asymmetric than for the other stars. 

With FAMIAS we fitted $\upsilon \sin i$ independent from known literature values. For all four stars we find higher values than current literature values, see Tables~\ref{propstar} and \ref{inppar}. The difference in the quoted values might be due to differences in equivalent width and $\xi_{\rm macro}$.
\newline
\newline
As discussed here, improvements in the spectral line profile analysis are still needed, both in terms of higher signal-to-noise ratio observations and in terms of the generated synthetic profiles we compare the data with. Expanding on an analysis of simulated data, as already started by \citet{hekker2006}, and applying the method to evolved stars with solar-like oscillations observed with high signal-to-noise ratio as we expect to become available from for instance SONG (Stellar Observations Network Group), will improve our understanding and increase the value of this method for mode identification in solar-like oscillators.

\begin{table}
\begin{minipage}{8.5cm}
\caption{Intervals for the surface rotation frequencies ($\Omega$) for the four evolved stars discussed here, together with the inclination angle ($i$) and projected rotational velocity ($\upsilon \sin i$), we used to compute $\Omega$.}
\label{srv}
\centering
\renewcommand{\footnoterule}{}
\begin{tabular}{lccc}
\hline\hline
star & $\Omega$ & $i$ & $\upsilon \sin i$\\
 & $\mu$Hz & degree & km\,s$^{-1}$\\
\hline
$\beta$~Hydri & 3.6 - 5.5 & 38 - 72 & 4.3\\
$\epsilon$~Ophiuchi & 0.8 - 1.1 & 45 - 81 & 5.7\\
$\eta$~Serpentis & 1.5 - 2.1 & 41 - 73 & 5.6\\
$\delta$~Eridani & 3.1 - 4.9 & 38 - 76 & 4.9\\
\hline
\end{tabular}
\end{minipage}
\end{table}

\begin{acknowledgement}
SH wants to thank Maarten Mooij for useful discussions and Wolfgang Zima for his help with FAMIAS. SH acknowledges financial support from the Belgian Federal Science Policy (ref: MO/33/018) and the UK Science and Technology Facilities Council. The research leading to these results has received funding from the European Research Council under the European Community's Seventh Framework Programme (FP7/2007--2013)/ERC grant agreement n$^\circ$227224 (PROSPERITY), as well as from the Research Council of K.U.Leuven grant agreement GOA/2008/04. We would like to thank our referee for valuable comments, which helped to improved the manuscript considerably.
\end{acknowledgement}

\bibliographystyle{aa}
\bibliography{bibpulsII}
\listofobjects
\end{document}